\documentclass[twocolumn]{aastex631}
\usepackage {graphicx}
\usepackage {amsmath}
\usepackage {bm}
\usepackage {comment}
\usepackage {textcomp}
\usepackage {mathcomp}
\usepackage[percent]{overpic}
\usepackage {float}
\usepackage {enumerate}
\usepackage {here}
\usepackage {color}

\newcommand{\Secref}[1]{Section~\ref{#1}}
\newcommand{\appref}[1]{Appendix~\ref{#1}}
\newcommand{\Tabref}[1]{Table~\ref{#1}}
\newcommand{\Eqref}[1]{Equation~(\ref{#1})}
\newcommand{\Figref}[1]{Figure~\ref{#1}}
\newcommand{\pr}[1]{\left({#1}\right)}

\renewcommand{\deg}[1]{{#1}^\circ}
\newcommand{\<}[1]{\langle{#1}\rangle}
\newcommand{\PDopt}{\<{\Pi_{\rm opt}}}
\newcommand{\PDradio}{\<{\Pi_{\rm radio}}}
\newcommand{\PAb}{\overline{\rm PA}_b}
\newcommand{\PAa}{\overline{\rm PA}_a}

\shorttitle{GRB afterglow polarization}
\shortauthors{Kuwata et al.}

\graphicspath{{./}}

\begin{document}

\title{Large-scale Magnetic Field Model of GRB Afterglow Polarization: Effects of Field Anisotropy, Off-axis Viewing Angle, and Ordered Field}

\correspondingauthor{Asuka Kuwata}
\email{a.kuwata@astr.tohoku.ac.jp}

\author[0000-0002-6169-2720]{Asuka Kuwata}
\affiliation{Astronomical Institute, Graduate School of Science, Tohoku University, Sendai 980-8578, Japan}

\author[0000-0002-7114-6010]{Kenji Toma}
\affiliation{Frontier Research Institute for Interdisciplinary Sciences, Tohoku University, Sendai 980-8578, Japan}
\affiliation{Astronomical Institute, Graduate School of Science, Tohoku University, Sendai 980-8578, Japan}

\author[0000-0001-7952-2474]{Sara Tomita}
\affiliation{Institute for Advanced Academic Research, Chiba University, 1-33 Yayoi-cho, Inage-ku, Chiba, 263-8522, Japan}

\author[0000-0003-3383-2279]{Jiro Shimoda}
\affiliation{Institute for Cosmic Ray Research, The University of Tokyo, 5-1-5 Kashiwanoha, Kashiwa, Chiba 277-8582, Japan}

\begin{abstract}
The afterglows of gamma-ray bursts are non-thermal electron synchrotron emissions from relativistic shocks. The origin of strong magnetic field in the emission region remains elusive, and two field amplification mechanisms via the plasma kinetic and magnetohydrodynamic instabilities have been discussed. The polarimetric observations are a powerful probe to distinguish these two mechanisms. So far, most theoretical works have focused on the former mechanism and constructed afterglow polarization models with microscopic-scale turbulence whose coherence length is much smaller than the thickness of the blast wave. In this work, focusing on the latter mechanism, we utilize our semi-analytic model of the synchrotron polarization with large-scale turbulence whose coherence length is comparable to the thickness of the blast wave to investigate the effect of magnetic field anisotropy and the observer viewing angle. We find that the polarization in our large-scale turbulence model can exhibit both behaviors characteristic of the microscopic-scale turbulence model and those not seen in the microscopic-scale model. Then we find that the large-scale model could explain all the polarimetric observational data to date that seem to be forward shock emission. We also examine the effect of ordered-field component, and find that polarization degree and polarization angle constant in time are realized only when the energy density ratio of the ordered and fluctuated components is $\gtrsim 50$. In this case, however, the polarization degree is much higher than the observed values.
\end{abstract}


\section{Introduction} \label{sec:intro}
Gamma-ray bursts (GRBs) are the brightest gamma-ray transients in the Universe. It is well understood that most GRBs are cosmological sources and they are powered by relativistic jets driven by a central engine. The central engine is believed to be a compact object, a black hole or a millisecond magnetar, formed by massive star collapses or binary compact star mergers \citep{Hjorth2003, Stanek2003, Hjorth2012, Abbott2017}. After the jet radiates the prompt gamma-ray emission, it sweeps up the external medium and is decelerated by a reverse shock, while a forward shock propagates into the external medium powering the long-lived afterglow emission \citep[see][for reviews]{Meszaros2002, Piran2004, Kumar2015}.\par

Afterglows in the broad energy range from radio to gamma-rays are synchrotron and inverse Compton radiation from non-thermal electrons, but the energy dissipation process at the relativistic collisionless shocks is still unclear. Comparison of the theoretical model and the observed multiwavelength light curves suggests that the turbulent magnetic field at the forward shock is amplified at least two orders of magnitude stronger than the shock-compressed magnetic field of the interstellar medium (ISM) \citep{Santana2014}. In close relation to the field amplification and the non-thermal particle acceleration, two turbulent magnetic field models have been discussed.\par

The first one is the plasma kinetic scale turbulence, which is induced by the Weibel instability \citep{Medvedev1999, Kato2005, Sironi2011, Ruyer2018, Takamoto2018, Lemoine2019}. This microscopic-scale turbulence is ubiquitous at the collisionless shock front. However, many kinetic simulations show that it decays substantially within 100 times of the proton plasma skin depth scale and it would not account for the observed synchrotron flux (\citealp{Gruzinov1999, Chang2008, Keshet2009, Tomita2016, Tomita2019}; but see \citealp{Groselj2024}). Moreover, the number fraction of non-thermal electrons to the thermal ones in the current particle-in-cell (PIC) simulations is smaller than that suggested by observations \citep{Sironi2011}. The other type of turbulent magnetic field is induced by magneto-hydrodynamic (MHD) instability, such as the Richtmyer-Meshkov instability \citep{Sironi2007, InoueT2013, Duffell2014, Mizuno2014}. Such a large-scale turbulence is favored for acceleration of electrons which emit observed GeV photons \citep{Asano2020, Ze-LinZhang2021, Huang2022}, but the conditions for large-scale field amplification are not clear in recent MHD and PIC simulations \citep{Tomita2022, Morikawa2024}. These studies provide us with some interesting hints, but the current numerical simulations have not examined all the relevant physical effects yet due to artificial particle mass ratios, limited simulation durations, and so on. Therefore, one cannot conclude which type of turbulence is dominant in the afterglow emission region only by comparing the current simulations and multi-wavelength light-curve data. In this work, we aim to solve the problem with help of polarimetric data.

Theoretical models of forward-shock afterglow synchrotron polarization have been constructed for the above two distinct scales of turbulence, and they predict different polarimetric features. In the microscopic-scale field model \footnote{
We define the `microscopic scale' and `large scale' as scales much smaller than and comparable to the thickness of the blast wave, respectively. They are referred to as `plasma scale' and `hydrodynamic scale', respectively, in \cite{Kuwata2023}.}, the synchrotron polarization from uniform jets shows characteristic behavior that the polarization degree (PD) takes the peak value and the polarization angle (PA) flips at $\deg{90}$ around the jet break time, when $\Gamma_{\rm sh} \simeq 1/\theta_j$, where $\theta_j$ is the jet opening angle \citep{Ghisellini1999, Sari1999}. The optical PDs are the same or higher than the radio PDs, and the PA difference between the optical band and radio band is $\deg{0}$ or $\deg{90}$ \citep{Rossi2004, Birenbaum2021, ST2021}. In the large-scale field model, \cite{Kuwata2023} (hereafter K23) constructed a semi-analytic model of turbulence and investigated the multi-band polarimetric feature in the case of the zero-viewing angle observer and the isotropic field. In this setup, the PDs and PAs are shown to change randomly and continuously, as expected by \cite{Gruzinov1999}. The optical PDs and the radio PDs are the same levels on average, which can be analytically estimated as $\sim 2f_B\%$, where $f_B$ is the ratio of the coherence length of turbulent magnetic field and the thickness of the blast wave (see \Eqref{eq:lambda-B} in \Secref{subsec:B-field}). Important is that the radio PDs can be higher than the optical PDs at some time intervals.\par

The optical polarimetric observations of forward shocks show that the typical measured PDs are $\sim 1\%-5\%$ \citep{Covino2016}. The temporal variabilities of PDs and PAs around the jet break time have been examined in four events, GRB 091018, GRB 121024A \citep{Wiersema2014}, GRB 020813 \citep{Gorosabel2004, Lazzati2004}, and GRB 191221B \citep{Urata2023}. The PA data of the former two events, especially GRB 121024A, are consistent with the microscopic-scale turbulence model, while those of the latter two events are difficult to explain by the microscopic-scale turbulence model even if considering the strong ordered magnetic field \citep{Granot2003, Lazzati2004} or the structured jet \citep{Rossi2004}. The number of events with measured radio polarization has increased recently \citep{Urata2019, Urata2023, Laskar2019}. The polarimetric observations of early optical afterglows are drawing attention \citep{Shrestha2022, Arimoto2023:2310.04144v1, Fernandez2024} and the radio polarimetric observation of gravitaional-wave associated GRB 170817A was conducted \citep{Corsi2018}. Such multi-band polarimetric observations of various types of GRBs with various timescales will help us distinguish the two turbulent field models.\par

\begin{deluxetable}{ll}
	\tablecaption{Parameters in this work \label{table: param}}
	\centering
	\tablehead{
    \colhead{Parameter} & \colhead{Value}
    }
    \startdata
		Redshift $z$ & $1.0$ \\
	    Isotropic energy of blast wave $E_{\rm iso}$ & $2.0\times 10^{52}\ {\rm erg}$  \\
		Upstream number density $n$ & $1.0\ {\rm cm^{-3}}$ \\
		Power-law index of accelerated electrons $p$ & $2.5$ \\
		Energy fraction of accelerated electrons $\epsilon_e$ & $0.1$ \\
		Energy fraction of magnetic field $\epsilon_B$ & $0.01$ 
	\enddata
\end{deluxetable}

K23 suggested that the large-scale field model can explain the observed polarimetric behaviors which are inconsistent with the microscopic-scale model. They only focused on the zero viewing angle observers and the isotropic turbulent magnetic field.
In this work, we investigate the dependence of polarization in our model on the field anisotropy, the viewing angle, and the ordered field. We also explore the compatibility of the model with the observational data.\par

This paper is organized as follows. In \Secref{sec:Kuwata23a}, we briefly summarize our formalism based on K23. In \Secref{sec:anisotropic-field}, \Secref{sec:off-axis} and \Secref{sec:ordered-field}, we show the numerical calculation results in the cases of the anisotropic magnetic field, the off-axis observer and the addition of ordered field, respectively. In \Secref{sec:PD-spectrum}, the spectra of PD and PA for the various cases are shown. Then, we discuss the compatibility of the large-scale field model and the microscopic-scale model with polarimetric observations in \Secref{sec:obs}. We summarize our findings in \Secref{sec:summary}.

\begin{deluxetable}{lcccccll}
	\tablecaption{Parameter sets in this work. The cases of $f_B \neq 1$ are also discussed in the text. \label{table: parameter-list}}
	\centering
	\tablehead{
        \colhead{Model}
        & \colhead{$\xi^2$} & \colhead{$\eta$} & \colhead{$f_B$} & \colhead{$\theta_j$} & \colhead{$\theta_v$} & \colhead{Explanation}
        }
        \startdata
        I1        & 1 & 0 & 1 & $\deg{4}$ & $\deg{2}$ & isotropic field\\
        A1        & 0 & 0 & 1 & $\deg{4}$ & $\deg{2}$ & anisotropic field\\
        I2        & 1 & 0 & 1 & $\deg{4}$ & $\deg{8}$ & isotropic field \& off-axis\\
        A2        & 0 & 0 & 1 & $\deg{4}$ & $\deg{8}$ & anisotropic field \& off-axis\\
        O         & 1 & 50 & 1 & $\deg{4}$ & $\deg{2}$ & ordered plus fluctuated field
	\enddata
\end{deluxetable}

\section{Synchrotron polarization of forward shocks with large-scale turbulent magnetic field} \label{sec:Kuwata23a}
\begin{figure}[t]
    \centering
    \includegraphics[width=1.0\linewidth]{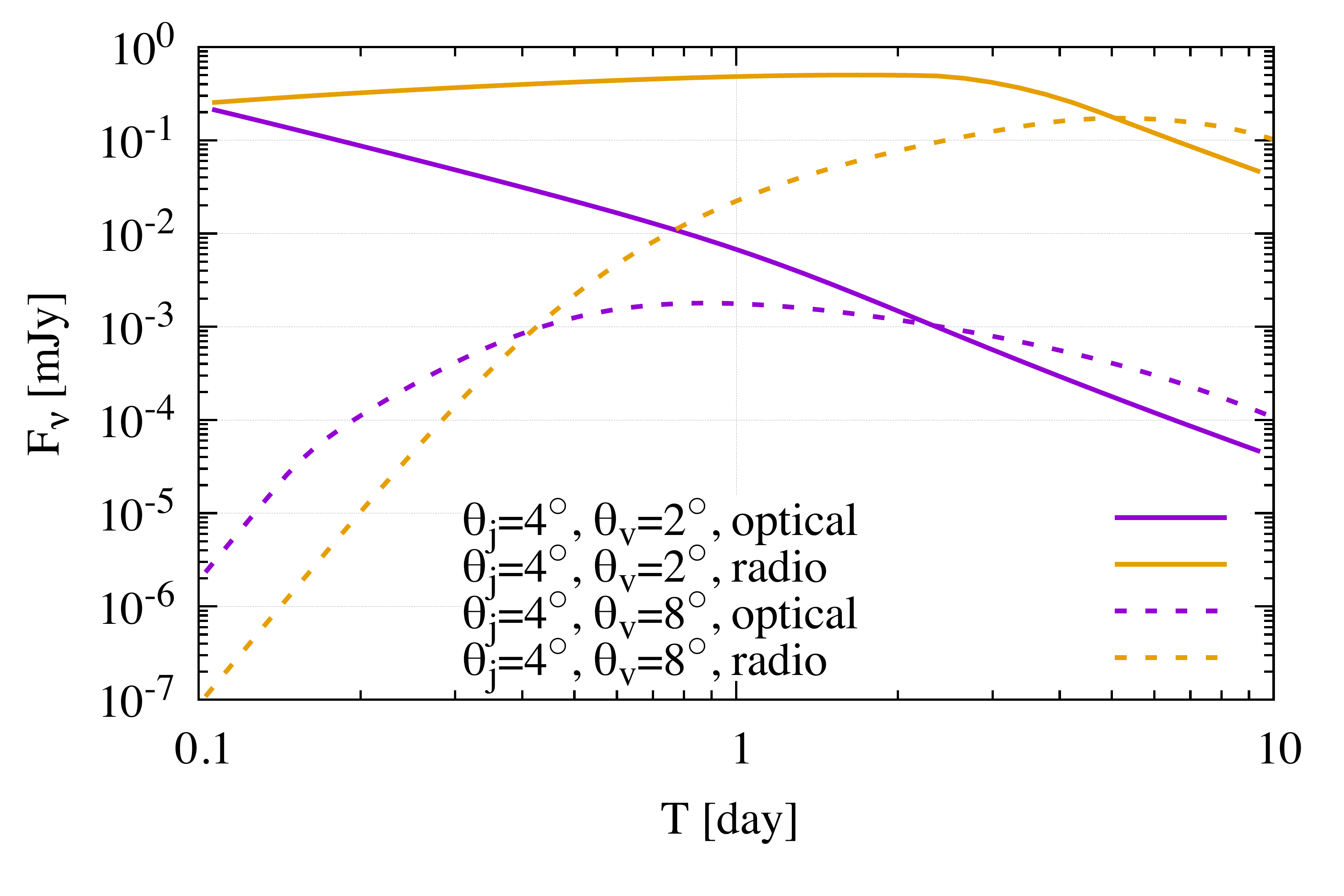}
    \caption{Light curves at frequencies $10^{15}$ Hz (optical, purple line) and 100 GHz (radio, orange line). The solid line corresponds to the on-axis observer (models I1 and A1), and the dashed line corresponds to the off-axis observer (models I2 and A2).}
    \label{fig:light-curve-model1&2}
\end{figure}

We calculate synchrotron polarization of forward shocks based on K23. The synchrotron light curves are calculated based on the standard model \citep{Sari1998, Granot1999a}. In Tables \ref{table: param} and \ref{table: parameter-list}, we show our model parameters used calculation of light curves, PDs, and PAs in this work. \Figref{fig:light-curve-model1&2} shows light curves in the case of the on-axis observer (models I1 and A1) and the off-axis observer (models I2 and A2). In our parameter sets, the optical frequency $\nu_{\rm opt}\; (10^{15}\; {\rm Hz})$ is at $\nu_m < \nu_{\rm opt} < \nu_c$ and the radio frequency $\nu_{\rm radio}\; (100\; {\rm GHz})$ is at $\nu_a < \nu_{\rm radio} < \nu_m$ for $T<T_{\nu_m}$, where $\nu_m$, $\nu_c$, $\nu_a$ and $T_{\nu_m}$ are the synchrotoron frequency of minimum electron Lorentz factor, the synchrotron cooling frequency, the self-absorption frequency, and the time when $\nu_m$ crosses $\nu_{\rm radio}$, respectively. The jet break time $T_{\rm jb}$ is $\sim 1\; {\rm day}$ and $T_{\nu_m}$ is $\sim 2\; {\rm day}$ for Models I1, A1. $T_{\nu_m}$ for Models I2, A2 is slightly later than that for Models I1, A1. We note that the approximation of the Blandford-McKee (BM) solution \citep{BM1976} used in this work is crude at $T = 4-10\;{\rm days}$ because $\gamma_l$ is $\sim 3-2$,\footnote{In the mildly-relativistic regime, the thickness of the blast wave is wider than in the relativistic regime, which will make PDs a bit lower than the calculation results with the BM solution (cf. K23). This effect should be examined in more details as a future work.} where $\gamma_l$ is the Lorentz factor of the fluid just behind the shock, which is the closest to the observer in the equal arrival time surface (EATS) \citep{Granot1999a}. We mainly examine the polarization at $T < 4\;{\rm day}$.

\begin{figure*}[htbp]
    \centering
    \includegraphics[width=1.0\linewidth]{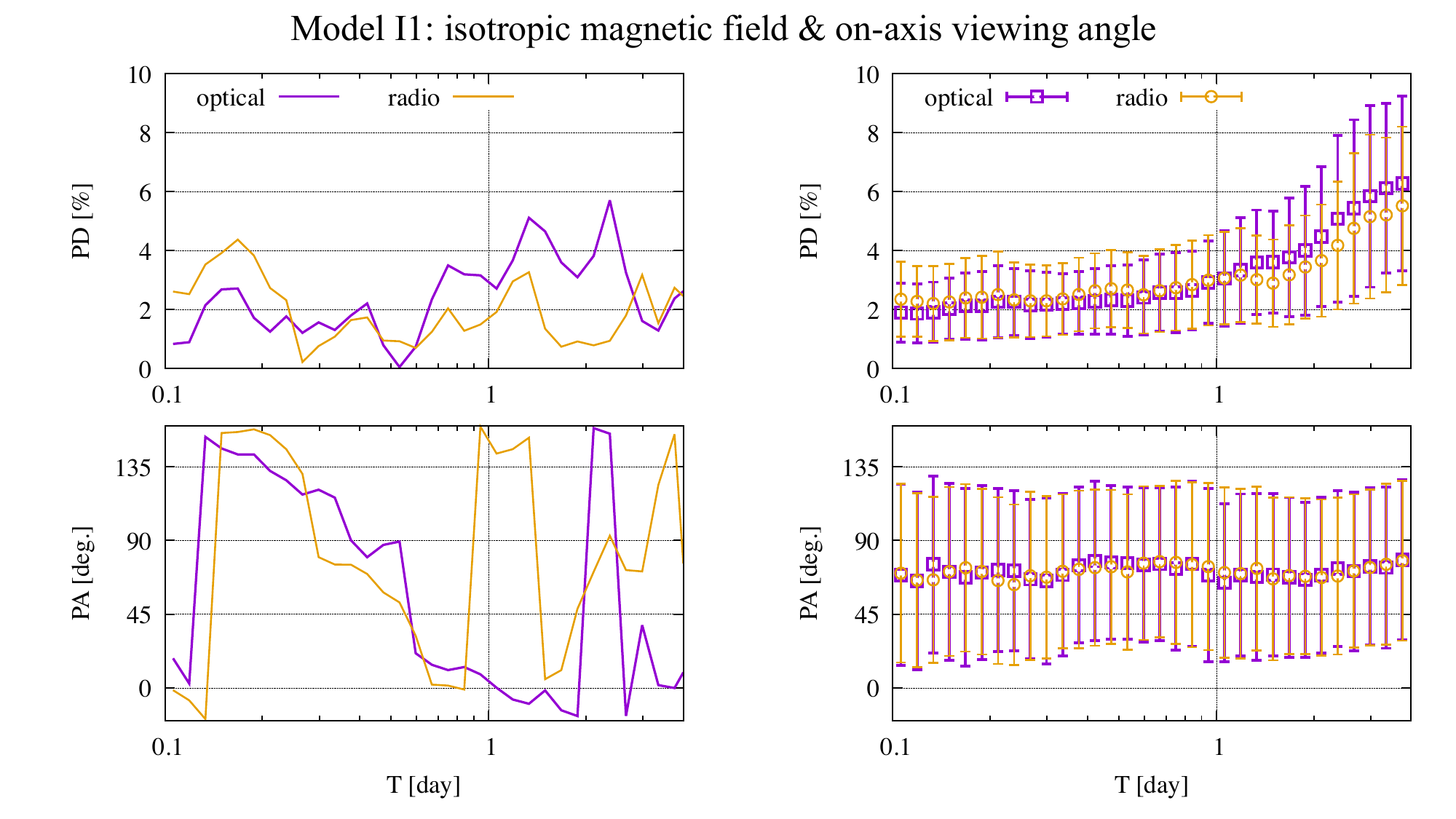}
    \caption{
    PDs and PAs as functions of time $T$ for Model I1 at $\nu_{\rm opt}$ (optical, purple line) and $\nu_{\rm radio}$ (radio, orange line) during $T = 0.1-4.0\;{\rm days}$. The top and bottom panels represent the PD curves, the PA curves respectively. (Left) PDs and PAs are calculated with one realization set of random numbers for the turbulence creation. (Right) Averaged PDs and PAs (the purple square for $\nu_{\rm opt}$ and the orange circle for $\nu_{\rm radio}$) and the standard deviations (the error bars) for 100 realization sets.  
    \label{fig:PDPA-curve-model1}}
\end{figure*}

\subsection{turbulent magnetic field model}\label{subsec:B-field}
We consider the combination of ordered magnetic field $\bm{B}_0$ and fluctuated turbulent magnetic field $\delta\bm{B}(x,y)$, $\bm{B}(x,y) = \bm{B}_0 + \delta\bm{B}(x,y)$.
We consider that the ordered field is simply shock-compressed magnetic field of the circumburst medium. The assumption on the fluctuated field component is set to be the same as in K23. We assume that the wavelength of fluctuated field is the order of the thickness of the blast wave $\sim \Delta R$ (see \Eqref{eq:lambda-B}). We also ignore the amplification timescale for simplicity. 

The fluctuated field component is defined as
\begin{equation}
    \delta\bm{B}(x,y) = \sum_{n=1}^{N_m} \bm{b}_n \exp{(i\bm{k}_n \cdot \bm{r} + i\beta_n)},
    \label{eq:B-field}
\end{equation}
where
\begin{equation}
    \label{eq:B-field-direction}
    \bm{b}_n = \sigma_{\perp} \cos \alpha_n \tilde{\bm{e}}_{y,n} + i\ \sigma_{\rm \|} \sin \alpha_n \bm{e}_z,
\end{equation}
the transformation from $\bm{r} = x\bm{e}_{x} + y\bm{e}_{y} + z\bm{e}_{z}$ to $\bm{r} = \tilde{x}_n\tilde{\bm{e}}_{x,n} + \tilde{y}_n\tilde{\bm{e}}_{y,n} + z\bm{e}_z$ is described by
\begin{equation}
	\begin{pmatrix}
		\tilde{x}_n\\
		\tilde{y}_n
	\end{pmatrix}
	=
	\begin{pmatrix}
		\cos \phi_n & \sin \phi_n  \\
		-\sin \phi_n & \cos \phi_n  \\
	\end{pmatrix}
	\begin{pmatrix}
		x\\
		y
	\end{pmatrix},
\end{equation}
and $\phi_n, \alpha_n, \beta_n$ are random numbers. We have two orthogonal directions $\tilde{\bm{e}}_{y,n}, \bm{e}_z$ of the turbulent magnetic field, which are chosen orthogonal to the wavevector $\bm{k}_n \| \tilde{\bm{e}}_{x,n}$ in order to satisfy $\bm{\nabla} \cdot \bm{B} = 0$ in the lab frame. For simplicity, the magnetic power spectrum is assumed to be concentrated at the scale of $k_0^{-1}$, i.e., we set $|\bm{k}_n| = k_0$. This makes $\sigma^2_{\perp}$ and $\sigma^2_{\parallel}$ independent of $n$, where $\sigma^2_{\perp}$ and $\sigma^2_{\parallel}$ are the variances of the wave amplitude in $\tilde{\bm{e}}_{y,n}$ and $\bm{e}_z$ directions, respectively.

We introduce three parameters for magnetic field configuration. The first one is the anisotropy of the fluctuated field in the comoving frame:
    \begin{equation}
        \xi^2 = \frac{2{\sigma'_\|}^2}{{\sigma'_\perp}^2},
    \end{equation}
so that we have $\xi^2 = 2 \gamma_f^2 \sigma_\|^2/\sigma_\perp^2$, where $\gamma_f$ is the Lorentz factor of the fluid just behind the shock of radius $R$. The second parameter is the ratio of the wavelength of turbulence to the thickness of the blast wave $f_B$. 
Let $\lambda'_B$ denote the wavelength corresponding to $k_0$ in the comoving frame. We assume that $\lambda'_B$ is of the order of $\Delta R' \simeq R/16\gamma_f$, and write
\begin{equation}
    \lambda'_{B} = f_B \frac{R}{16\gamma_f}.
    \label{eq:lambda-B}
\end{equation}
In the following sections, we show the figures of calculation results for $f_B=1.0$, and briefly discuss the results for $f_B \neq 1.0$ in the text.
The third parameter is the energy density ratio of the ordered-field to fluctuated-field components in the comoving frame:
    \begin{equation}
        \eta = \frac{\bm{B}^{\prime 2}_0}{\langle \delta \bm{B}^{\prime 2}(x,y) \rangle},
    \end{equation}
where $\langle \delta \bm{B}^{\prime 2}(x,y) \rangle = \sigma^{\prime 2}_\perp + \sigma^{\prime 2}_\|$.\par

\begin{figure*}[htbp]
    \includegraphics[width=1.0\linewidth]{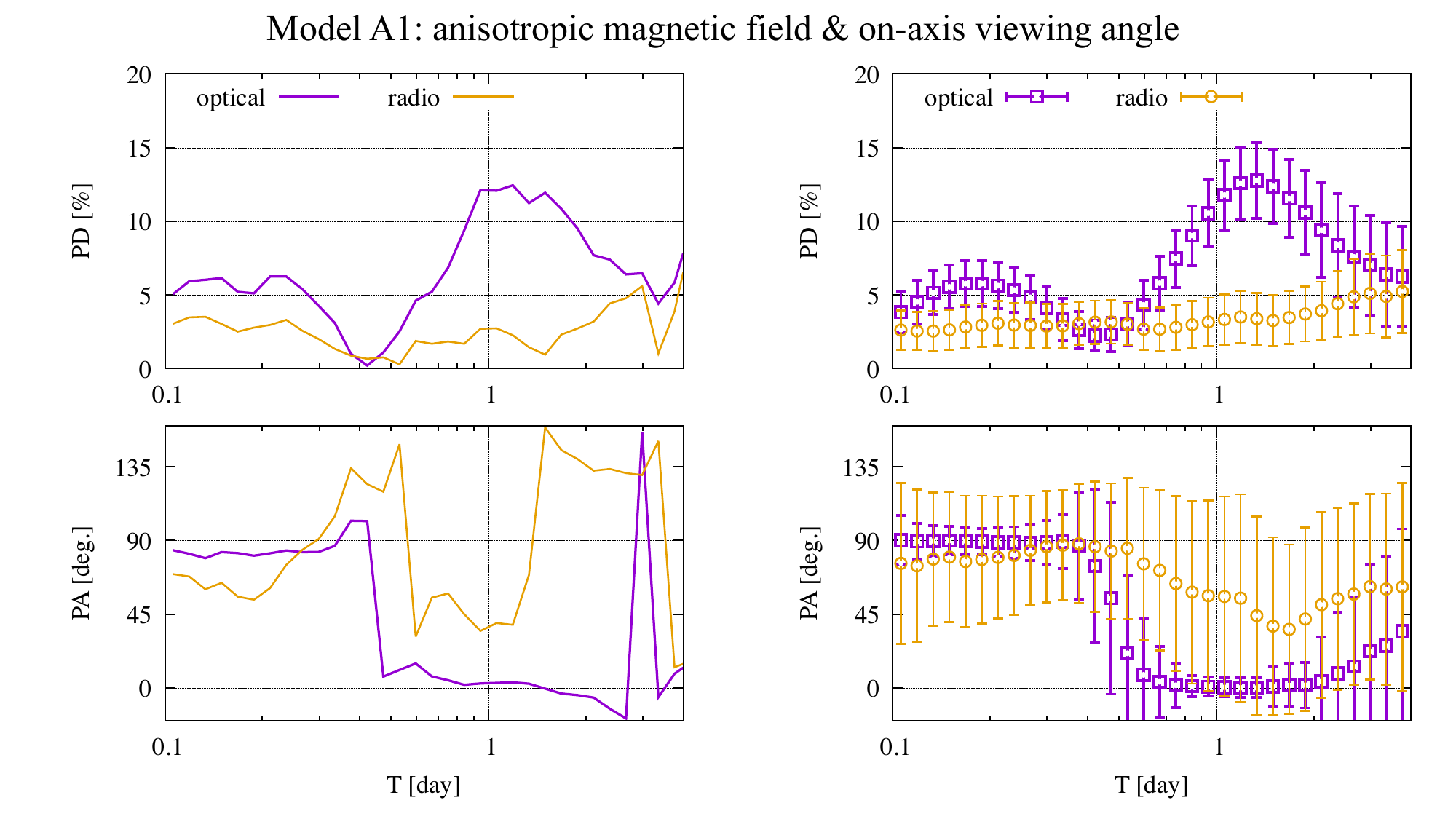}
    \caption{
    Same as \Figref{fig:PDPA-curve-model1}, but for Model A1.
    \label{fig:PDPA-curve-model4}}
\end{figure*}

\subsection{synchrotron polarization}\label{subsec:polari}
We calculate local PAs according to Eq.(26)-(28) in K23. We obtain observed Stokes parameters $I_\nu, Q_\nu, U_\nu$, which are integrated over the EATS according to \cite{Granot1999a}:
\begin{align}
	&I_\nu = \int \frac{dF_\nu}{dyd\phi} \sin^\epsilon \theta_B' dyd\phi, \label{eq:I-nu} \\
	&Q_\nu = \int \Pi_0 \cos(2\phi_p) \frac{dF_\nu}{dyd\phi} \sin^\epsilon \theta_B' dyd\phi, \label{eq:Q-nu} \\
	&U_\nu = \int \Pi_0 \sin(2\phi_p) \frac{dF_\nu}{dyd\phi} \sin^\epsilon \theta_B' dyd\phi, \label{eq:U-nu}
\end{align}
where $F_\nu$ is the flux density of radiation, $y$ and $\phi$ are coordinates on the EATS, $\phi_p$ is the PA at each grid of EATS, and $\epsilon = (p+1)/2$.
$\Pi_0$ is the synchrotron PD for the uniform magnetic field, which has frequency dependence \citep{Radipro,Melrose1980b},
\begin{equation}
	\label{eq:Pi-0}
	\Pi_0 = 
	\begin{cases}
		0.5 & (\nu < \nu_m) \\
		\frac{p+1}{p+7/3} & (\nu_m \leq \nu < \nu_c) \\
            \frac{p+2}{p+10/3} & (\nu \geq \nu_c)
	\end{cases}.
\end{equation}
Note that we focus on the spectral segments at $\nu_a < \nu < \nu_c$ (see Section 2.4 in K23). 
We obtain the net PD $\Pi_\nu$ and the net PA $\Phi_{p,\nu}$ using $I_\nu, Q_\nu, U_\nu$:
\begin{align}
	&\Pi_\nu = \frac{\sqrt{Q_\nu^2 + U_\nu^2}}{I_\nu}, \label{eq:pi-tot} \\
 	&\tan{(2\Phi_{p,\nu})} = \frac{U_\nu}{Q_\nu}. \label{eq:phi-tot}
\end{align}
We write the net PDs at $\nu_{\rm opt}$ and $\nu_{\rm radio}$ by $\Pi_{\rm opt}$ and $\Pi_{\rm radio}$, respectively.\par
In the following \Secref{sec:anisotropic-field}-\ref{sec:ordered-field}, we show our numerically calculated polarization with different parameter sets shown in \Tabref{table: parameter-list}.

\section{dependence on $\xi^2$} \label{sec:anisotropic-field}
\begin{figure*}[htbp]
    \centering
    \includegraphics[width=0.46\linewidth]{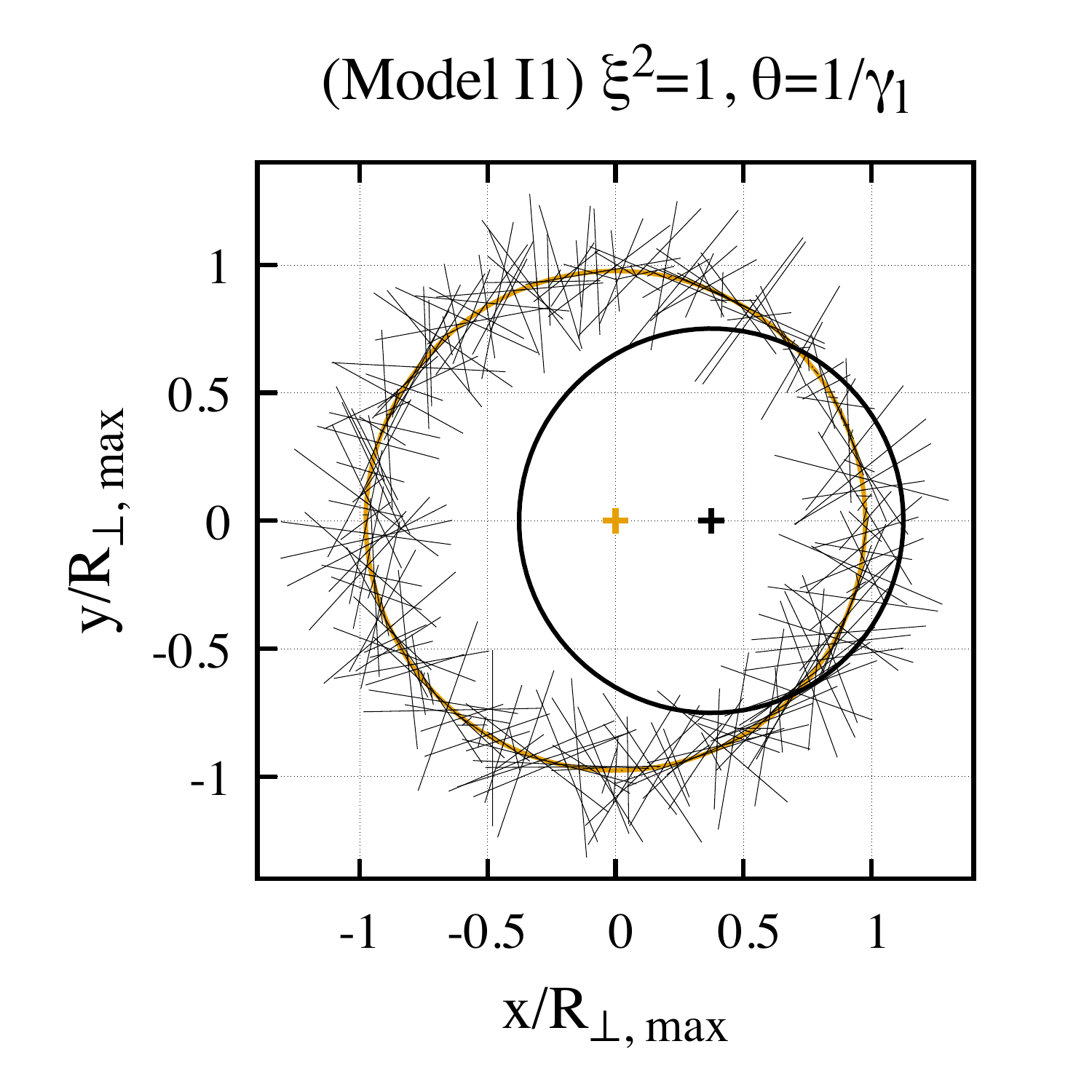}
    \includegraphics[width=0.46\linewidth]{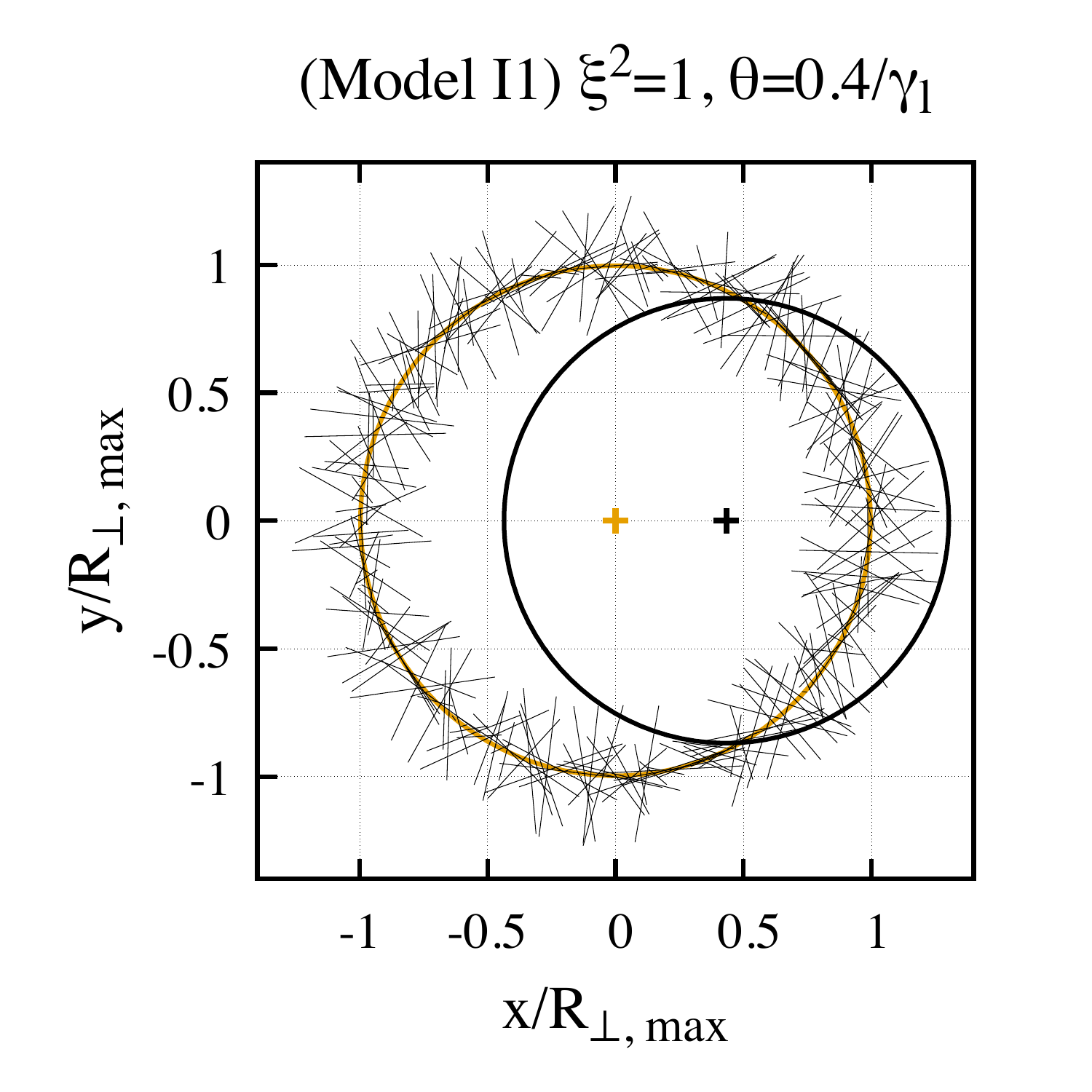}\\
    \includegraphics[width=0.46\linewidth]{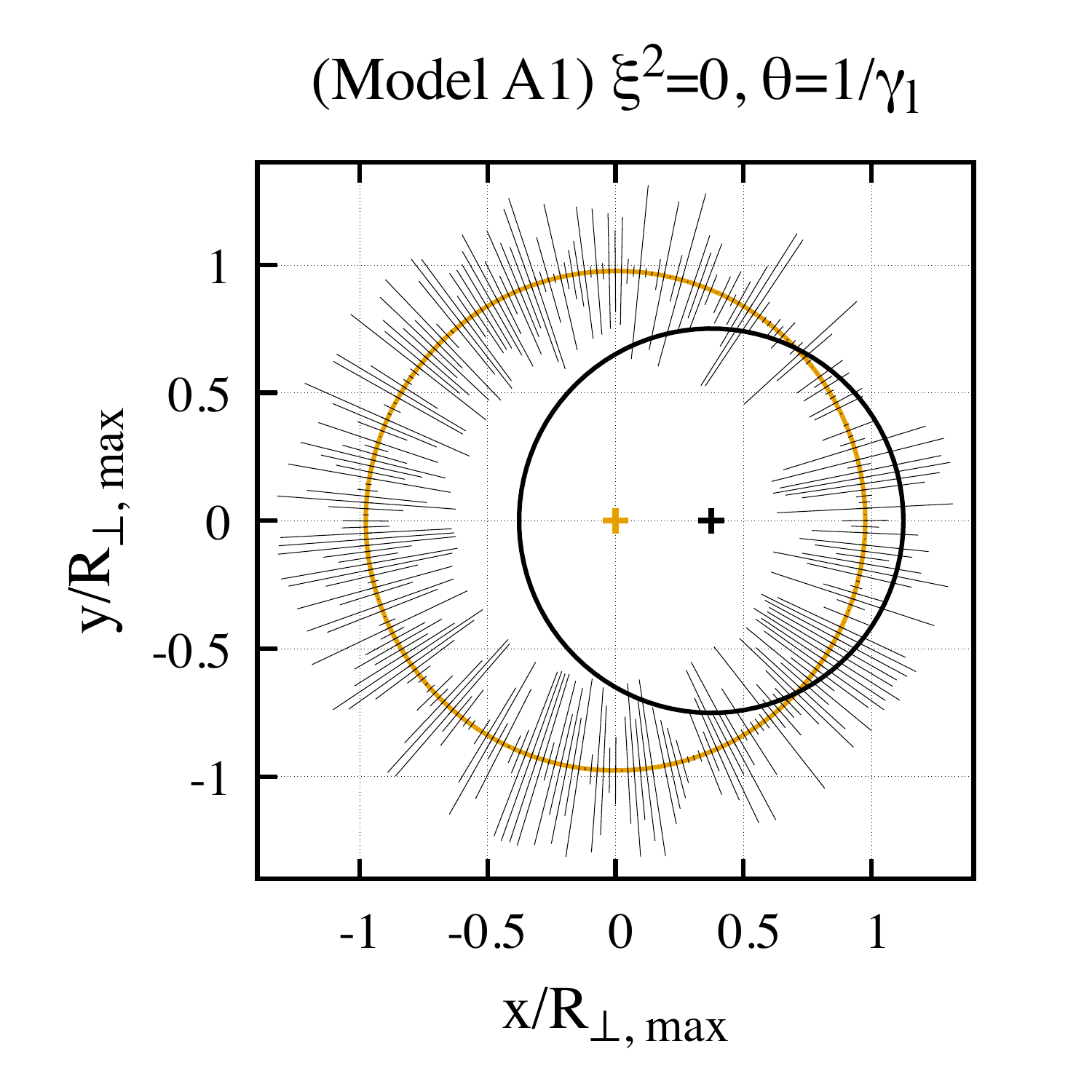}
    \includegraphics[width=0.46\linewidth]{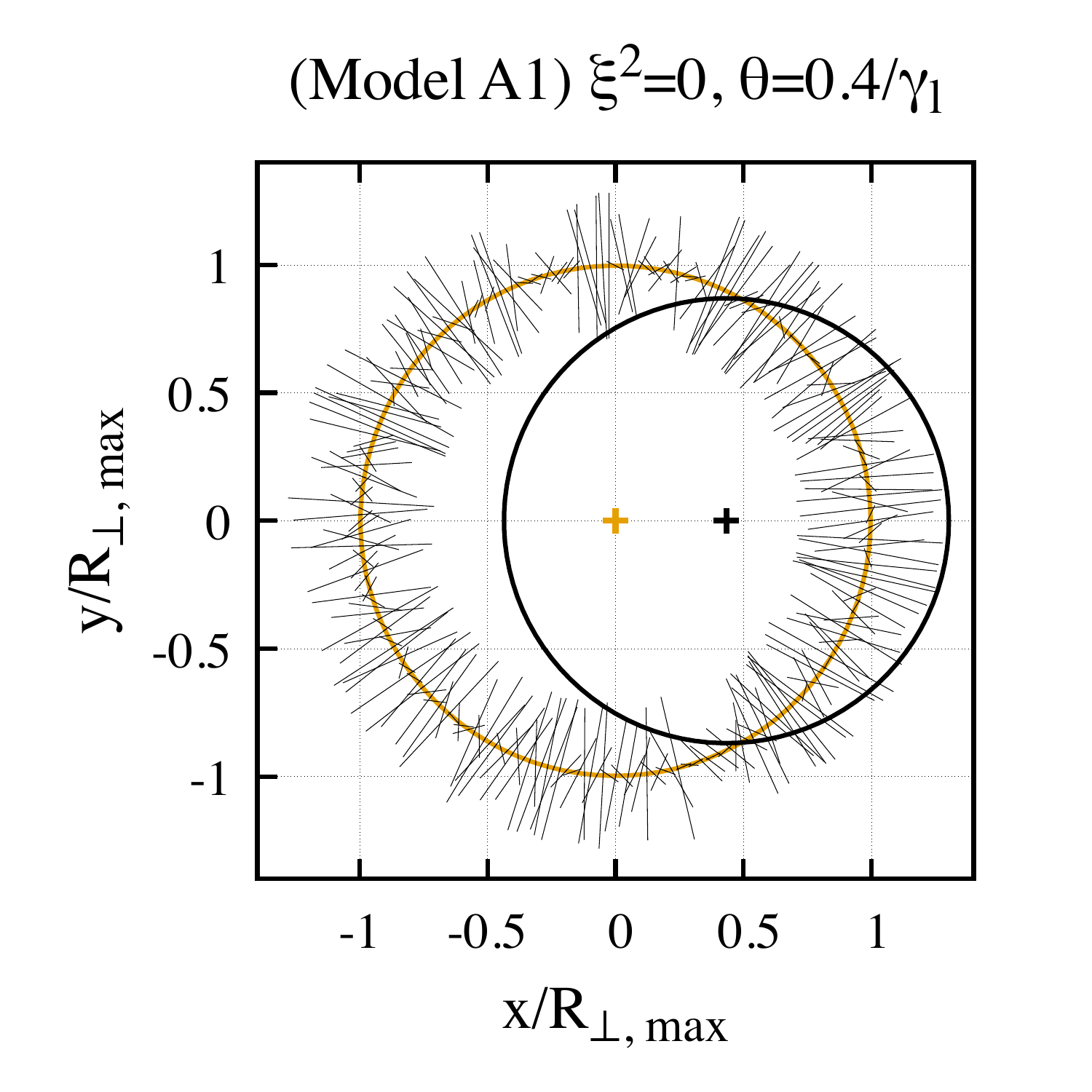}
    \caption{
    Local PA $\phi_p$ and the normalized polarized intensity $PI_\nu$ in Model I1 (upper) and A1 (bottom). The direction and length of each of the thin black bars indicate $\phi_p$ and the value of normalized $PI_\nu$ at the EATS with $\theta=1/\gamma_l$ (left) and $\theta=0.4/\gamma_l$ (right). The jet axis marked with a black "+" symbol and the line of sight is marked with an orange "+" symbol. We show just for reference the thin black bars outside the jet (thick black circle) from which no emission comes.
    \label{fig:local-PAPI}}
\end{figure*}

We show the dependence of the polarization on the field anisotropy $\xi^2$ in the case without ordered field ($\eta=0$) and the observer with the viewing angle $0<\theta_v<\theta_j$. We explain the calculation results of Model I1 ($\xi^2 = 1$, isotropic field) and Model A1 ($\xi^2 = 0$, highly anisotropic field) in detail and also summarize the characteristic features for other values of $\xi^2$.\par
\Figref{fig:PDPA-curve-model1} shows the PD and PA curves of Model I1 ($\xi^2 = 1$) at the frequencies of $\nu_{\rm opt}$ and $\nu_{\rm radio}$ during $T=0.1-4.0\;$days. The left side panels of \Figref{fig:PDPA-curve-model1} shows PDs and PAs calculated with one realization set of random numbers for the turbulence creation. The right side panels show the averaged PDs (denoted by $\PDopt$ and $\PDradio$) and PAs and the standard deviations for 100 realization sets of random numbers. This model represents the isotropic fluctuated field and has the same parameter set as K23 except $\theta_v$ and $\theta_j$. K23 focuses on the zero-viewing angle observer and sets $\theta_j=\deg{6}$, $\theta_v=\deg{0}$. The behaviors of the PD and PA curves for one realization and the realization-averaged PD and PA are consistent with K23. For reference, the results for other turbulence realizations in Model I1  (and in the other models of \Tabref{table: parameter-list} as well) are shown in \appref{app:realization}.\par

Model A1 is the anisotropic fluctuated field ($\xi^2 = 0$). \Figref{fig:PDPA-curve-model4} shows the calculated PD and PA curves of Model A1. The panels are displayed in the same manner as \Figref{fig:PDPA-curve-model1}. Interestingly, the behavior of the realization-averaged optical PD and PA curves is similar to that in the microscopic-scale turbulence model, in which PDs have one or two peaks, and PAs keep constant or have a sudden flip by $\deg{90}$ \citep{Ghisellini1999, Sari1999, Rossi2004}. This behavior can be seen also in the optical PD and PA curves calculated with the one realization set (left panels). On the other hand, the radio PD and PA curves show random behavior, which is similar to that in Model I1. $\PDradio$ is lower than $\PDopt$ when $\PDopt$ takes peak. On the other hand, $\PDradio$ is slightly higher than $\PDopt$ between two peaks of $\PDopt$. This feature is different from one of the microscopic-scale field model, in which the radio PD is always lower than the optical one.\par

We interpret the behavior of Model I1 and A1 from the distribution of local PAs. \Figref{fig:local-PAPI} shows the local PA $\phi_p$ and local polarized-intensity, 
\begin{equation}
    PI_\nu \equiv \sqrt{\pr{\frac{dQ_\nu}{dyd\phi}}^2+\pr{\frac{dU_\nu}{dyd\phi}}^2} =\Pi_0 \frac{dF}{dyd\phi} \sin^\epsilon \theta_B', \label{eq:local-PI}
\end{equation}
at $\theta = 1/\gamma_l$ and $\theta = 0.4/\gamma_l$, where $\theta$ is the angle measured from the line of sight, respectively. $\theta = 0.4/\gamma_l$ corresponds to $R_\perp = R_{\perp,\rm max}$, where $R_\perp$ is the distance of a position on the EATS from the line of sight and $R_{\perp, \rm max}$ is the maximum value of $R_\perp$, respectively. In model I1, $\phi_p$ changes randomly on the wavelength scale ($\lambda'_B \sim R_\perp/10$ for $f_B=1$) both at $\theta = 1/\gamma_l$ and at $\theta = 0.4/\gamma_l$. In Model A1, in contrast, $\phi_p$ at $\theta = 1/\gamma_l$ is radially oriented and ones at $\theta = 0.4/\gamma_l$ are also more radially oriented than Model I1. This is because in Model A1 both of the comoving photon direction $\bm{\hat{k^\prime}}$ and the comoving magnetic field direction $\bm{\hat{B^\prime}}$ are in the plane perpendicular to the shock normal, which makes the polarization direction $\bm{\hat{k^\prime}} \times \bm{\hat{B^\prime}}$ along the shock normal in the comoving frame, i.e., in the radial direction in the lab frame. Therefore, in Model I1, the net PD is determined by $\Pi \sim 70/\sqrt{N} \% \sim 2f_B \%$, where $N$ is the number of patches within which the magnetic field is coherent in the bright region (in the same manner as Sec.~4 of K23), while in Model A1, the radially oriented local PAs make the net PD higher than the one in Model I1 and the time variability of PDs look like the one of microscopic-scale field model.\par 

\begin{figure*}[htbp]
    \centering
    \includegraphics[width=0.9\linewidth]{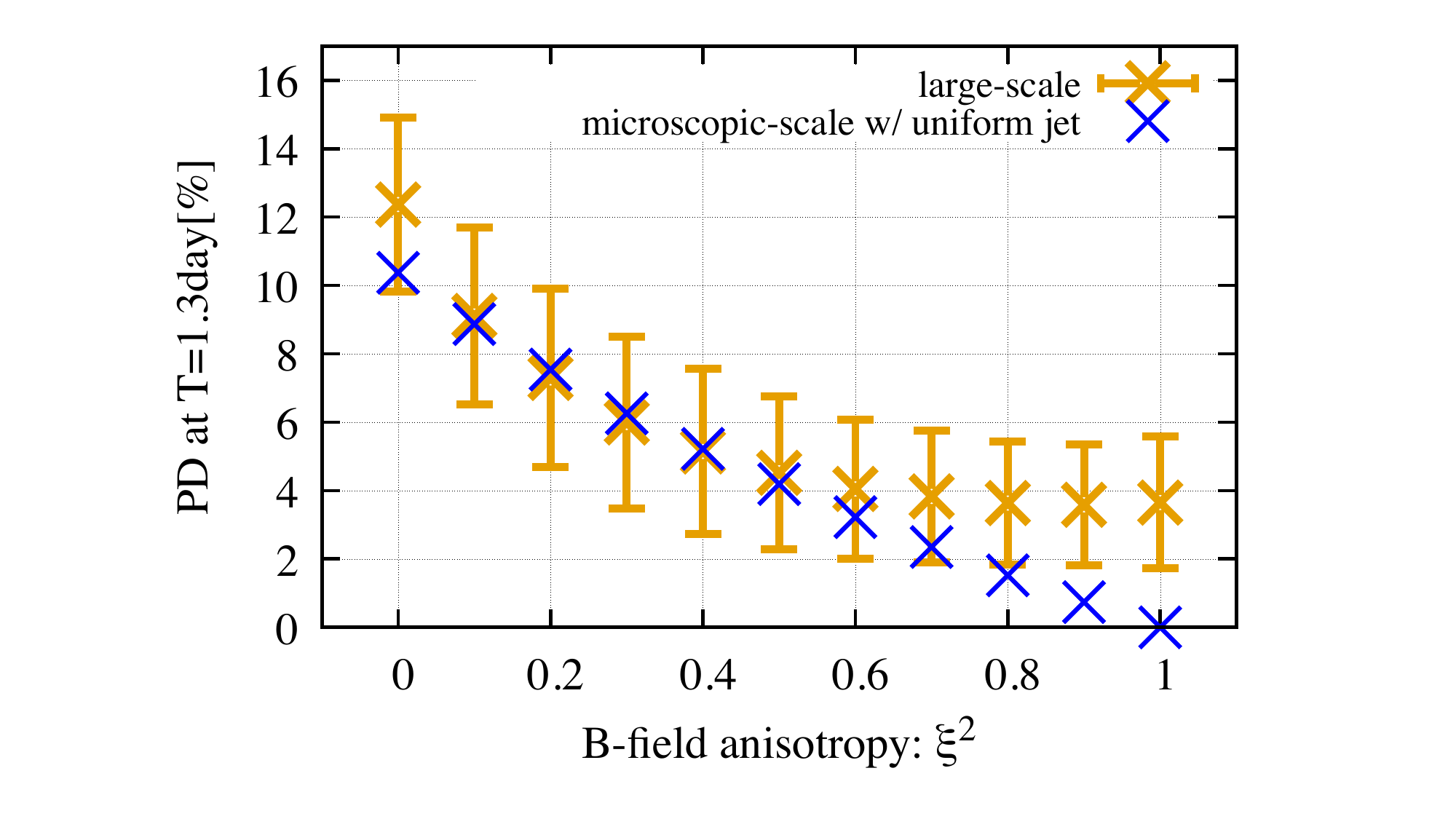}
    \caption{
    Dependence of the optical PD at $T=1.3\;$day on $\xi^2$. The orange marks show the averaged PDs over 100 turbulent realizations and their variances in the large-scale turbulence model with $f_B=1.0$. The blue marks show PDs in the microscopic turbulence model with uniform jets.
    \label{fig:xi-dependence-peakPD}}
\end{figure*}
\begin{figure*}[htbp]
    \centering
    \includegraphics[width=0.9\linewidth]{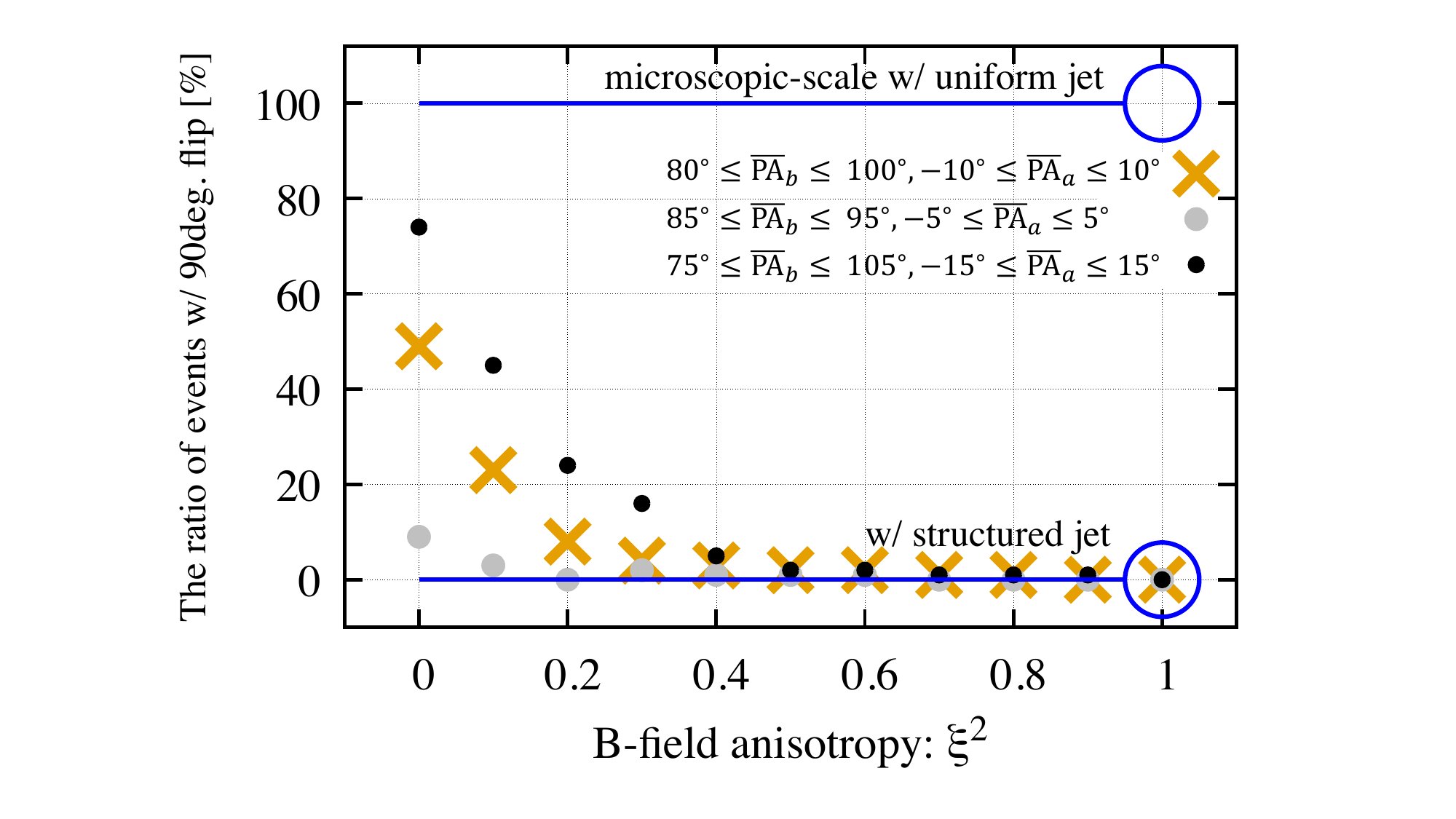}
    \caption{ 
    Dependence of the PA flip ratio on $\xi^2$.
    The orange, grey, and black marks represent the flip ratios in the large-scale field model with $f_B = 1.0$ for different criteria of flip. The two horizontal blue lines show the flip ratio in the microscopic-scale field model with uniform jets and that with structured jets, for which the PDs are always $0\%$ if $\xi^2=1$.
    \label{fig:xi-dependence-90flip}}
\end{figure*}

We show the dependence of the optical polarization on $\xi^2$ between $\xi^2=0$ and $1$ both in the large-scale model and the microscopic-scale model in \Figref{fig:xi-dependence-peakPD} and \ref{fig:xi-dependence-90flip}. We note that the following discussion holds even when $\xi^2>1$, by replacing $\xi^2$ with $\xi^{-2}$. \Figref{fig:xi-dependence-peakPD} shows the dependence of the optical PD on $\xi^2$ at $T=1.3\; {\rm day}$, the peak time of the optical PDs except for $\xi^2=1$ (or jet break time). The orange crosses and error bars represent $\PDopt$ and the standard deviations in the large-scale model with $f_B=1.0$. The blue crosses show the maximum optical PD in the microscopic-scale model. In \Figref{fig:xi-dependence-peakPD}, both of the models show the similar maximum PDs for $\xi^2 < 1$, for which PDs are zero in the microscopic-scale model while the net PDs $\sim 2f_B\%$ in the large-scale model (K23).\par
 
We also examine the dependence of the ratio of events which show $\deg{90}$ PA flips (hereafter `flip ratio') on $\xi^2$. In the microscopic-scale model with uniform jets, PAs before $T \sim 0.4\;$day are maintained at $\deg{90}$ and after that PAs flip to $\deg{0}$ if the parameter sets are the same in \Tabref{table: param}. In Model A1, similarly, the PA flip occurs between $\sim 0.4\;\rm{day}$ and $\sim 0.5\;\rm{day}$ (see \Figref{fig:PDPA-curve-model4}). To evaluate the PA flip, we define $\PAb$ and $\PAa$ as temporally averaged PA over $T=0.1-0.4\;$day and that over $T=0.5-4\;$day, respectively. We evaluate the flip ratio by the ratio of events which show $\deg{80} \leq \PAb \leq \deg{100}$ and $\deg{-10} \leq \PAa \leq \deg{10}$. In \Figref{fig:xi-dependence-90flip} we show the result, where we also plot the results with looser and more strict criteria for the PA flip. In the microscopic-scale model, the flip ratios are 100\% for the uniform jets \citep{Ghisellini1999, Sari1999, ST2021} and 0\% for the structured jets \citep{Rossi2004}, respectively. We note that for $\xi^2=1$ the PDs are always 0\% and the flip ratio is meaningless.\par

Here we briefly note the dependence on $f_B$. We calculate PD and PA curves in the case of $\xi^2 = 0, \theta_j = 4^\circ, \theta_v = 2^\circ$, and $f_B = 4.0$. This is the same parameter set as Model A1 except $f_B$. We find that the optical PDs at the first and second peak are almost the same values as the Model A1 ($f_B=1.0$). The optical PD between the two peaks and the radio PD before the jet break time, on the other hand, can be explained by $\sim 2 f_B \%$ for $f_B=4.0$. The optical PA shows $\deg{90}$ flip more gradually than the one of Model A1.

\section{Off-axis viewing angle} \label{sec:off-axis}
\begin{figure*}[htbp]
    \centering
    \includegraphics[width=1.0\linewidth]{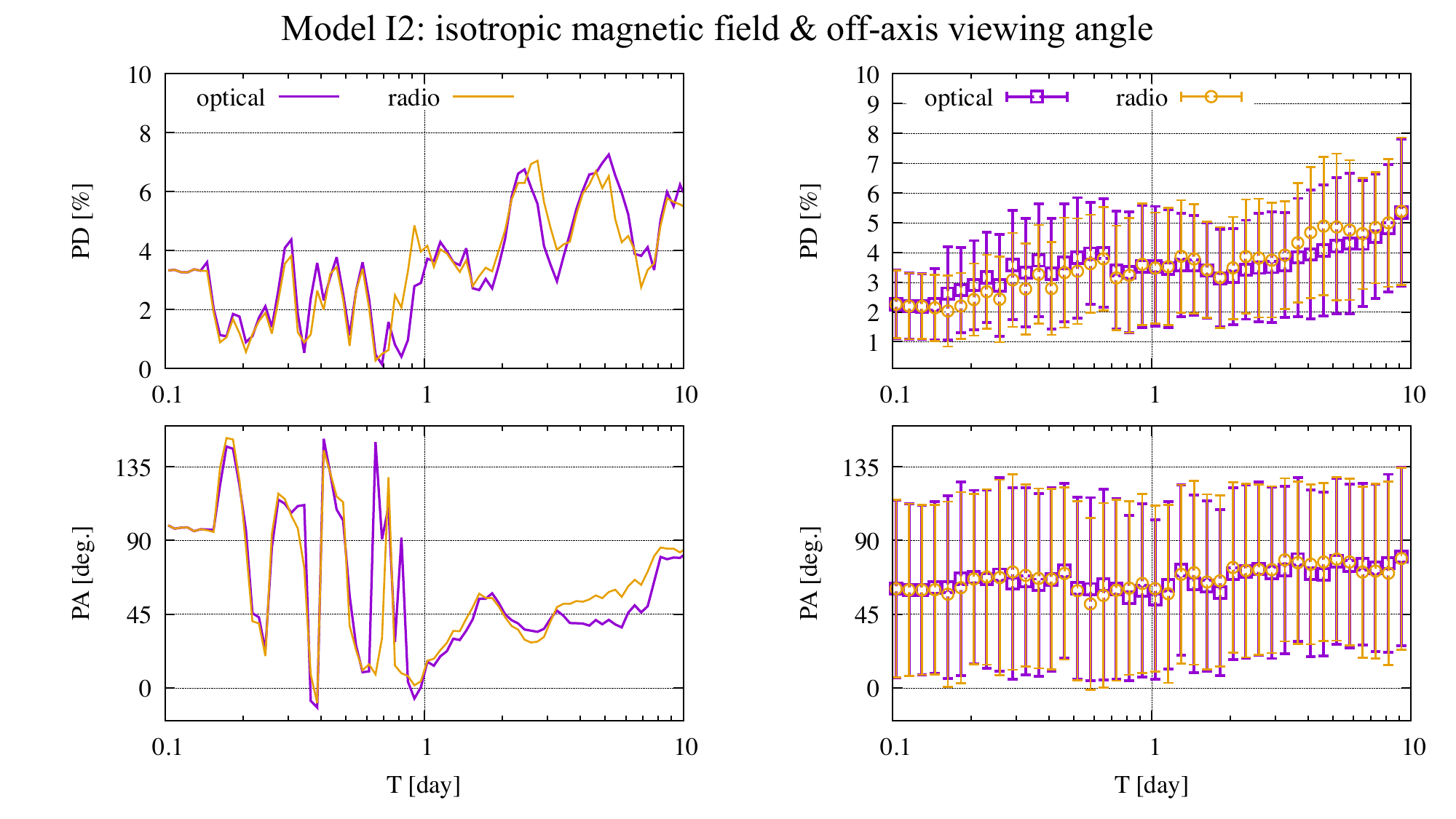}
    \caption{
    Same as \Figref{fig:PDPA-curve-model1}, but for Model I2. 
    \label{fig:PDPA-curve-modelI2}}
\end{figure*}

We show the result in the case of the off-axis observer ($\theta_v>\theta_j$). \Figref{fig:PDPA-curve-modelI2} (left) shows the PD and PA curves in the model of isotropic field and off-axis viewing angle (model I2). Both the optical PDs and PAs and the radio PDs and PAs show random behavior similar to those in Model I1. \Figref{fig:PDPA-curve-modelI2} (right) shows the realization-averaged PD and PA curves for Model I2. They are also similar to Model I1, except for the slightly higher $\PDopt \sim \PDradio \sim 4\%$ than that in Model I1. This is because $N$ for the off-axis observer is smaller than those for the on-axis observer. \par
\begin{figure*}[htbp]
    \centering
    \includegraphics[width=1.0\linewidth]{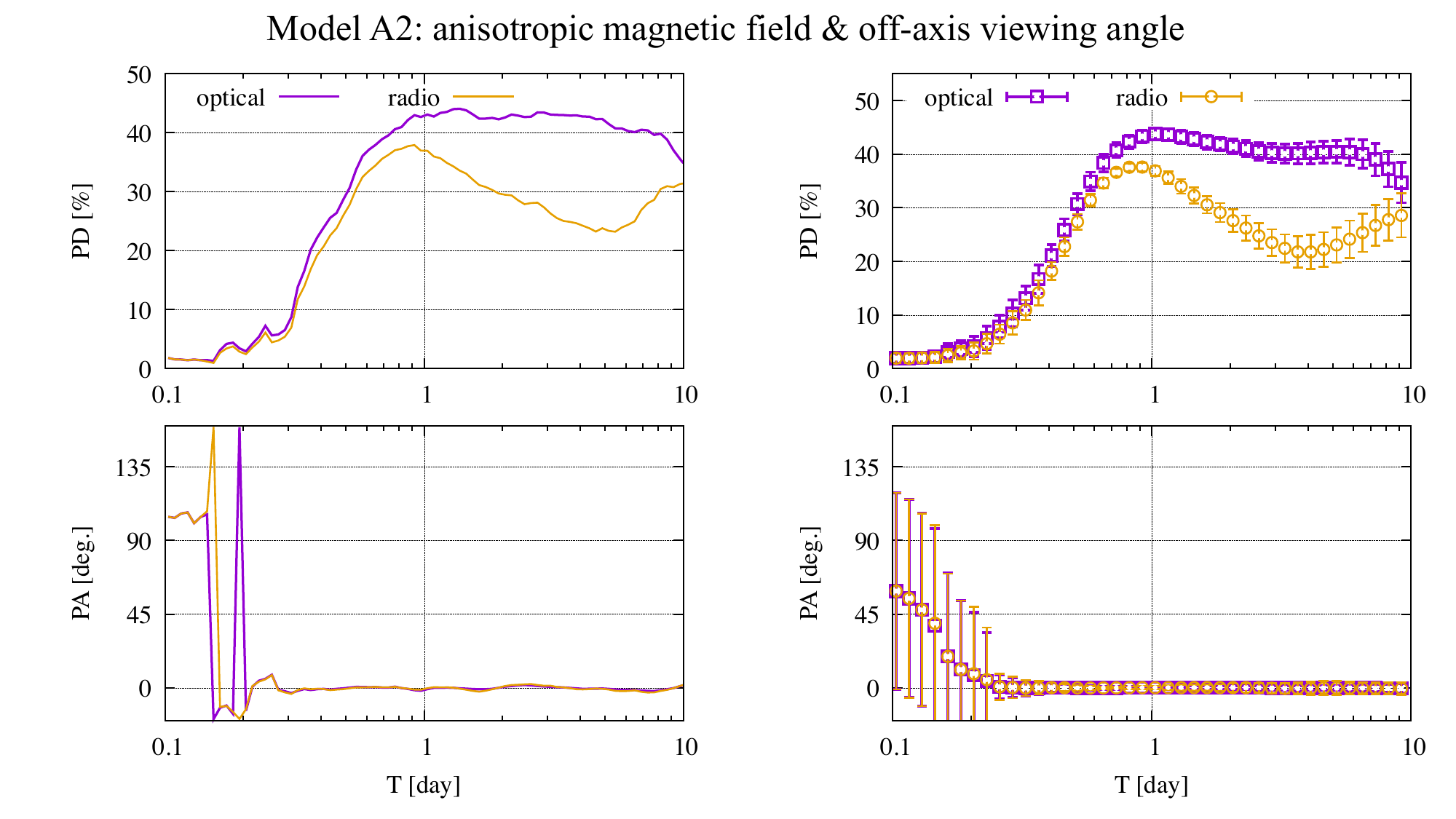}
    \caption{
    Same as \Figref{fig:PDPA-curve-model1}, but for Model A2.
    \label{fig:PDPA-curve-modelA2} }
\end{figure*}

\Figref{fig:PDPA-curve-modelA2} (left) shows the PD and PA curves for Model A2, the same as Model I2 except for field anisotropy. The optical PD maintains $\sim 40\%$ from $\sim 1\;\rm{day}$ to $\sim 10\;\rm{day}$ and the optical PA is almost constant after $T\sim0.3\;\rm{day}$, which are similar behavior to the optical PD and PA in the microscopic-scale model with the off-axis observer \citep[see Figure 8 of][]{Rossi2004}. The optical PA at $T \lesssim 0.3\;\rm{day}$ changes randomly, reflecting the randomness of the local PA. The radio PD peaks at $\sim 1\;\rm{day}$ and increases again after $\sim 3\;\rm{day}$, which is due to $\nu_m$ crossing at $\nu_{\rm radio}$ \citep[cf.][]{ST2021}. The radio PA also keeps constant after $T\sim0.3\;\rm{day}$ similarly as 
the optical one. The behaviors of realization-averaged PD and PA shown in the right panel of \Figref{fig:PDPA-curve-modelA2} are almost the same as the example shown in the left panel.
\par

\begin{figure*}[htbp]
    \centering
    \includegraphics[width=1.0\linewidth]{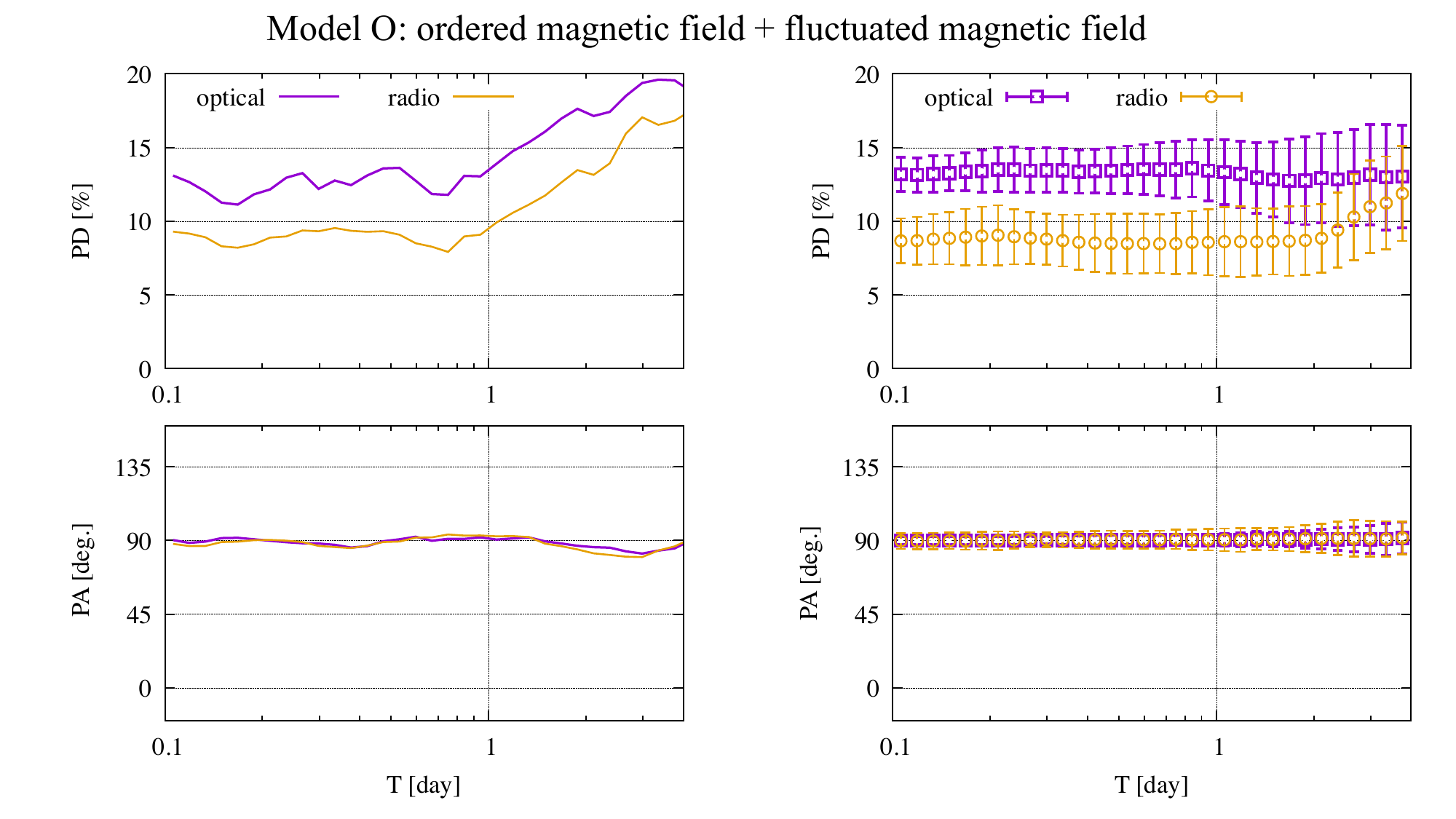}
    \caption{
        Same as \Figref{fig:PDPA-curve-model1}, but for Model O. 
        \label{fig:PDPA-curve-modelO}}
\end{figure*}

\section{Contribution of ordered magnetic field} \label{sec:ordered-field}
In this section, we show the effect of ordered magnetic field on polarization. We consider that the ordered magnetic field is along $x$-axis in the lab frame, $\bm{B}_0 \parallel \bm{x}$. \Figref{fig:PDPA-curve-modelO} (left) shows the PD and PA curves in model O ($\eta=50$). One can see that the optical and radio PAs are almost constant, which means that the contribution of ordered field on polarization is dominant for $\eta \gtrsim 50$. Both the optical PD and the radio PD are higher than those in Model I1. The optical PD is always higher than the radio PD, which is due to the difference of $\Pi_0$ (see \Eqref{eq:Pi-0}). \Figref{fig:PDPA-curve-modelO} (right) shows the realization-averaged PD and PA curves for Model O. The realization-averaged optical PD and radio PD are constant values of $\PDopt \sim 13\%$ and $\PDradio \sim 9\%$ at $T\lesssim T_{\nu_m}$, which are much higher than observed PDs to date (see \Secref{sec:obs}). The realization-averaged PAs are also constant similar to the example shown in the left panel.\par

We investigate the value of $\eta$ for which the effect of the ordered-field appears on the polarization. For $\eta \sim 1$, the difference of optical PDs from Model I1 is up to $\sim 0.5\%$, whereas the radio PDs are not much different. For $\eta \sim 3$, the difference of optical PDs is up to $\sim 1\%$, and the difference of radio PDs is up to $\sim 0.5\%$. Therefore, the effect of ordered field can appear on polarization at $\eta \gtrsim 1$ in the optical band and at $\eta \gtrsim 3$ in the radio band.\par

We estimate the required strength of ordered field when the effect of the ordered field appears on the polarization. From the definition of $\epsilon_B$, $|\bm{B}^{\prime}_0| + \sqrt{\langle \delta \bm{B}^{\prime 2}(x,y) \rangle} = \sqrt{32\pi \epsilon_B \gamma_f^2 n_0 m_{\rm p} c^2}$, we estimate the ordered-field strength as 
\begin{equation}
    \pr{1+\frac{1}{\sqrt{\eta}}}|\bm{B}^{\prime}_0| = 0.2\;{\rm G}\pr{\frac{\epsilon_B}{10^{-2}}}^{\frac{1}{2}} \pr{\frac{\gamma_f}{5}} \pr{\frac{n_0}{1\;{\rm cm^{-3}}}}^{\frac{1}{2}}.
\end{equation}
When $\eta \sim 1$, the upstream ordered-field strength in the lab frame is $|\bm{B}_{0, \rm up}| = |\bm{B}^\prime_0|/4\gamma_f = 5\;{\rm mG}$ for $\gamma_f \sim 5 [ (E_{\rm iso}/2\times 10^{52}\; {\rm erg}) (n/1\;{\rm cm^{-3}})^{-1} (T/1\;{\rm day})^{-3} ((1+z)/2)^{3} ]^{1/8}$. This is over three orders of magnitude stronger than the typical magnetic field strength of the ISM ($\sim 1\;\mu {\rm G}$). Therefore, in the ISM cases, the effect of the ordered field will not appear on the polarization.\par

In the case of the progenitor star wind model, the typical ordered-field (toroidal field) strength at the afterglow emission region is
$ |\bm{B}_{\rm AG}| \sim |\bm{B}_{*}| \pr{ R/R_* }^{-1}
                \approx 0.1\;{\rm mG} \pr{ |\bm{B}_{*}|/1\;{\rm kG} }
                    \pr{ R/5\times10^{17}\;{\rm cm} }^{-1}
                      \pr{ R_*/1R_\odot }, $
where $|\bm{B}_{*}|$ is the typical magnetic field strength of the surface of Wolf-Rayet (WR) stars, $\sim 1\;{\rm kG}$ \citep{Hubrig2016}, $R \sim 5.5\times 10^{17}\; {\rm cm}\; [ (E_{\rm iso}/2\times 10^{52}\; {\rm erg}) (n/1\;{\rm cm^{-3}})^{-1} (T/1\;{\rm day}) ((1+z)/2)^{-1} ]^{1/4}$, and $R_*$ is the typical radius of WR stars, $\sim R_\odot$. When $\epsilon_B$ is reasonably small $\sim 10^{-5}$, $|\bm{B}_{0, \rm up}|$ is comparable to $|\bm{B}_{\rm AG}|$. This comparison suggests that the effect of the ordered field appears on polarization in the wind model.\par

In the microscopic-scale model, \cite{Teboul&Shaviv2021} calculate the polarization by summing up the random and the ordered field, and the effect of the ordered field appears at $\eta \gtrsim 0.3$. \cite{Granot2003} also calculate the polarization in the presence of ordered field by summing the Stokes parameters, and find that the effect of the ordered field appears at $\eta > P_{\rm rnd}/P_{\rm ord}$, where $P_{\rm rnd}\sim 2\%$ is PD from the random field only and $P_{\rm ord} \sim 75\%$ is PD from the ordered field only. However, the linear combination of Stokes parameters is only valid when the emission region of the ordered field and that of the random field are different.\par

\section{PD and PA spectra} \label{sec:PD-spectrum}
\begin{figure}[htbp]
        \centering
        \includegraphics[width=0.9\linewidth]{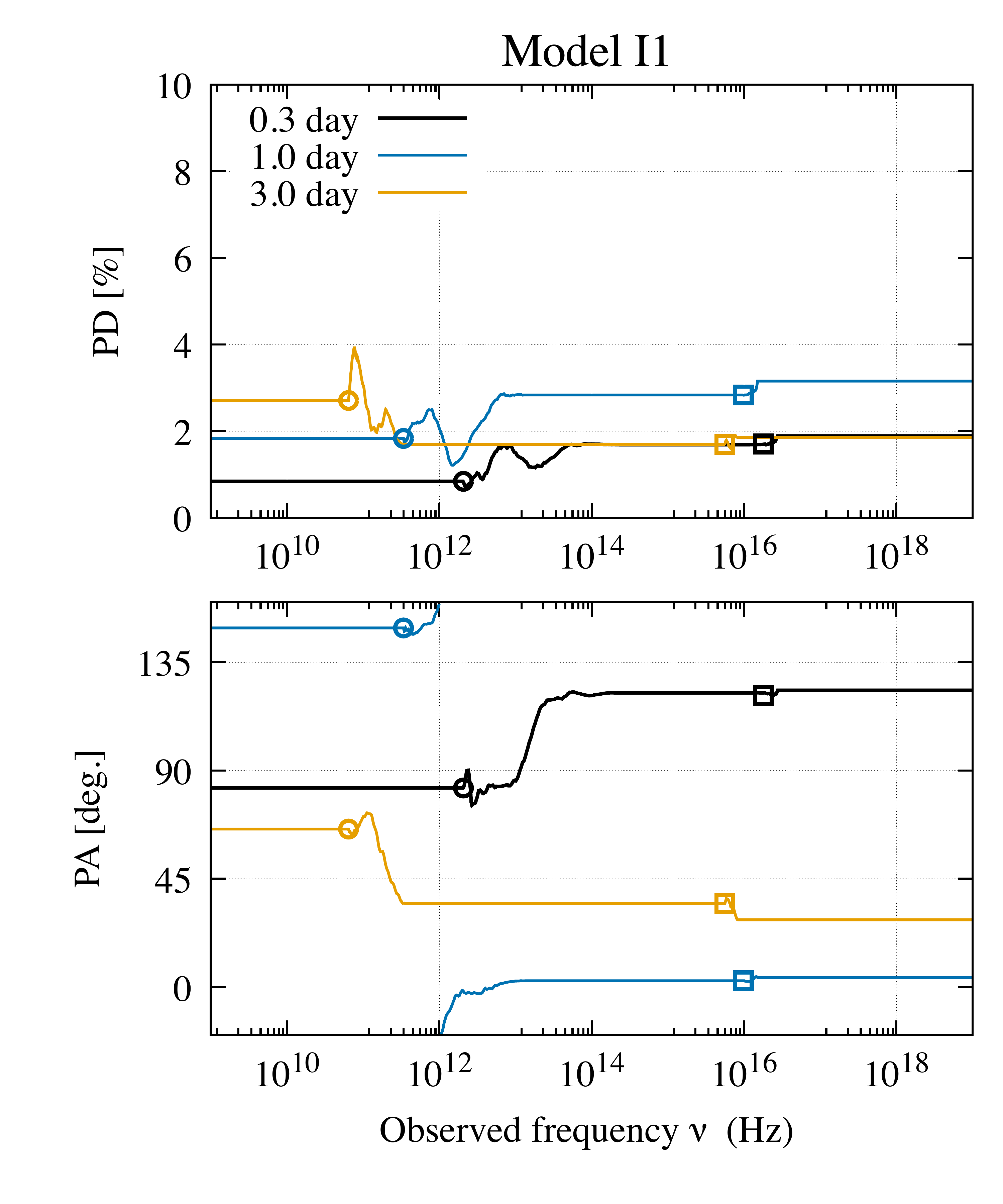}
        \caption{
        PDs and PAs as functions of the observed frequency $\nu$ for Model I1 at $T=0.3(<T_{\rm j})$, $1(\sim T_{\rm j})$ and $3(\sim T_{\rm \nu_m})\;{\rm days}$. The circle points and square points indicate $\nu_{\rm m}$ and $\nu_{\rm c}$, respectively. 
        \label{fig:PDPA-spectrum-modelI1}}
\end{figure}

\begin{figure}[htbp]
        \centering
        \includegraphics[width=0.9\linewidth]{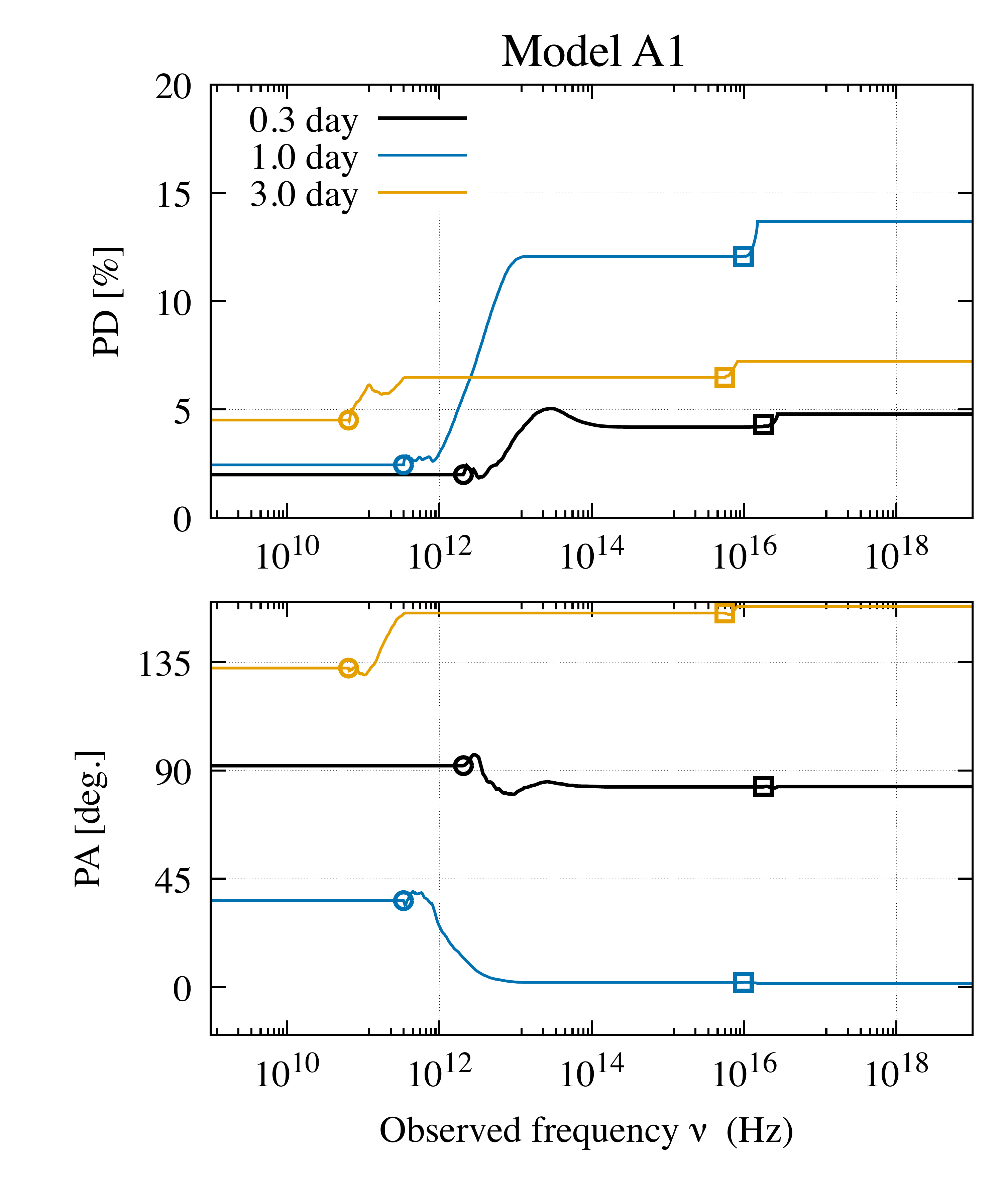}
        \caption{
        Same as \Figref{fig:PDPA-spectrum-modelI1}, but for Model A1. 
        \label{fig:PDPA-spectrum-model4}}
\end{figure}

\begin{figure}[htbp]
        \centering
        \includegraphics[width=0.9\linewidth]{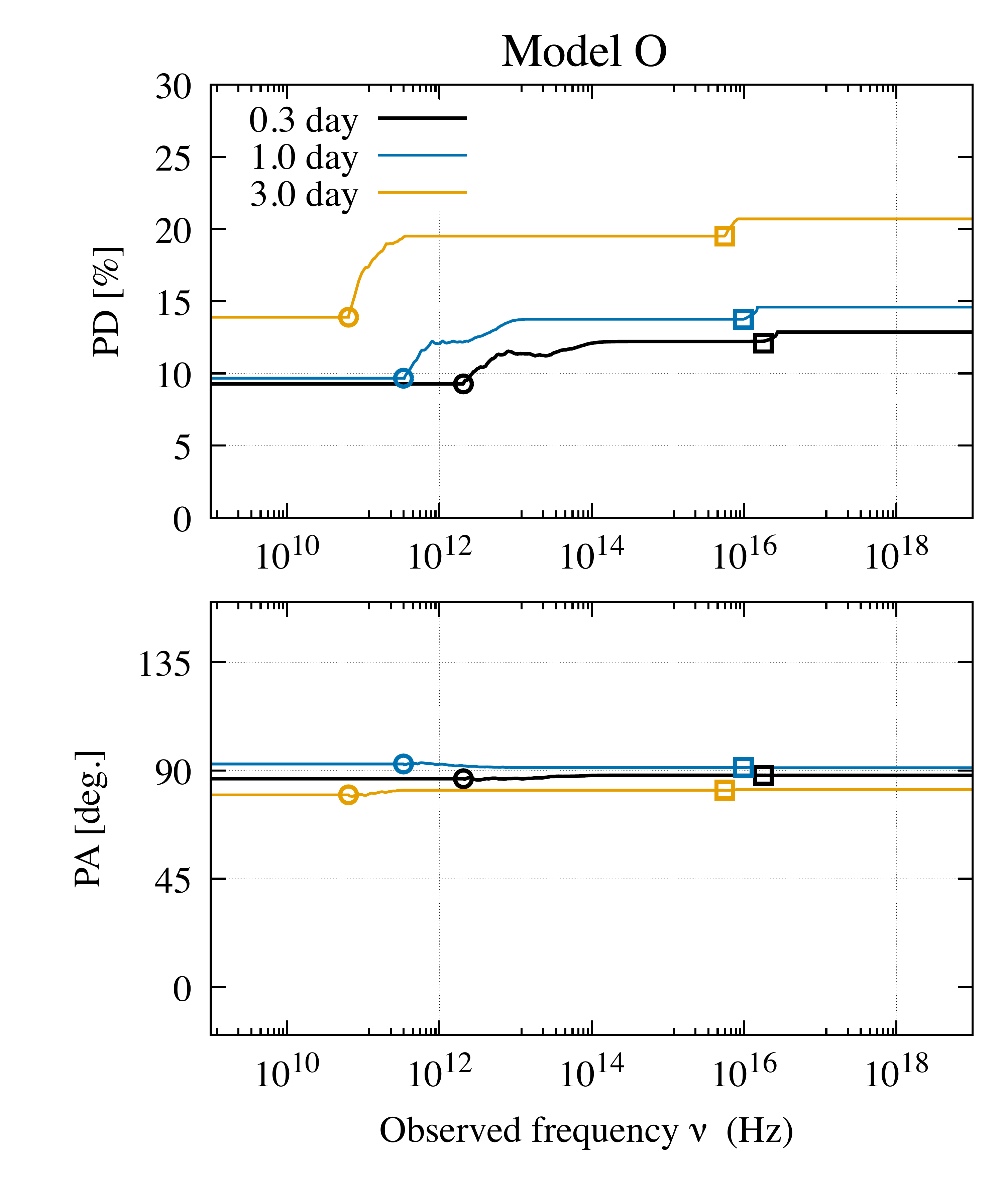}
        \caption{
        Same as \Figref{fig:PDPA-spectrum-modelI1}, but for Model O. 
        \label{fig:PDPA-spectrum-modelO}}
\end{figure}

\Figref{fig:PDPA-spectrum-modelI1}, \ref{fig:PDPA-spectrum-model4}, and \ref{fig:PDPA-spectrum-modelO} show the PD spectra in Model I1, A1, and O, respectively. We show the PD spectra at $T=0.3\;(<T_{\rm j})$, $1\;(\sim T_{\rm j})$ and $3\;(\sim T_{\rm \nu_m})\;$days with one turbulence realization. We also calculate PD spectra with two different realizations. In mode I1, \Figref{fig:PDPA-spectrum-modelI1} shows that the optical PD is higher than the radio one during $T\leq T_{\rm j}$, while with the other two different realizations, the optical PD is lower than the radio one at the same time interval, which is not seen in the microscopic-scale model. The difference between the optical PA and the radio PA takes various values at each time and realization. The PD and PA takes constant value at $\nu \ll \nu_m, \nu_m \ll \nu \ll \nu_c, \nu_c \ll \nu$ because $\Pi_0$ in \Eqref{eq:Pi-0} and the surface brightness profile do not vary at each of these frequency segments \citep[c.f.][]{Granot1999a}. We note that the PDs as well as PAs are similar in the optical and the X-ray bands, because of the similarity of the surface brightness for these bands.\par


If a fraction of electrons in the shock downstream are just isotropized, not heated to the level of protons' temperature, the Faraday depolarization effect makes PD and PA strongly depend on frequencies \citep{Toma2008}. This effect is tiny in the microscopic-scale turbulent field.\par

In Model A1 (\Figref{fig:PDPA-spectrum-model4}), the optical PD is always higher or equal to the radio one, which is the same frequency dependence in the microscopic-scale model. The PA difference between the optical and the radio takes various value. On the other hand, in the microscopic-scale model, the optical PA is the same as or differs from the radio one by $\deg{90}$. Therefore, we can distinguish between the microscopic-scale and the large-scale models from the PA difference between the optical and the radio.\par
In Model O (\Figref{fig:PDPA-spectrum-modelO}), the X-ray PDs are always higher than the optical PDs, and the optical PDs are always higher than the radio PDs, reflecting the frequency dependence of $\Pi_0$ in \Eqref{eq:Pi-0}. The PAs are $\deg{90}$ regardless of frequency, reflecting the dominance of ordered field.

\section{Discussion: comparison to observational data}
\label{sec:obs}

We have shown in \Secref{sec:anisotropic-field}-\ref{sec:PD-spectrum} that the polarimetric feature in the large-scale turbulence model depends on $\xi^2, \theta_v/\theta_j,$ and $\eta$. In the case of the isotropic field (Model I1, \Figref{fig:PDPA-curve-model1}), both PD and PA show random behavior in time and the average PD is $\sim 2f_B\%$ at $T<T_{\rm jb}$. When the magnetic field is anisotropic (Model A1, \Figref{fig:PDPA-curve-model4}), on the other hand, the optical PD takes a peak at $T\sim T_{\rm jb}$, and the optical PA shows a $\deg{90}$ flip, which is similar to the behavior in the microscopic-scale model. The maximum level of PD and the flip ratio depend on $\xi^2$ (\Figref{fig:xi-dependence-peakPD} and \ref{fig:xi-dependence-90flip}). In the case of the off-axis observer, when the magnetic field is isotropic (Model I2, \Figref{fig:PDPA-curve-modelI2}), PD and PA show time evolution similar to that in the case of the on-axis observer. In the case of off-axis observer and anisotropic field (Model A2, \Figref{fig:PDPA-curve-modelA2}), the temporal behavior of PD and PA is similar to that in the microsopic-scale model. We note that in the case of the off-axis observer, the PDs tend to be higher independently of the field anisotropy. The effect of the ordered-field component on polarization (Model O, \Figref{fig:PDPA-curve-modelO}) appears when $\eta \gtrsim 1$ in the optical band and $\eta \gtrsim 3$ in the radio band. The PD and PA are constant in time only when $\eta \gtrsim 50$, in which case PD is much higher than the observed value.\par
In this section, we discuss how the polarimetric data obtained so far can be explained with the large-scale field model. The microscopic-scale model is also discussed. The discussions are summarized in \Tabref{tab:obs-summary}.

\subsection{observation of late afterglows}\label{subsec:obs-late-afterglow}
\textbf{Typical polarimetric data of late optical afterglows}: 
The typical optical PDs measured at $T \gtrsim 10^4\;{\rm sec}$ range for $\sim 1-5\%$ \citep{Covino2004, Covino2016} \footnote{We note that GRB 020405 shows the optical PD of $9.89 \pm 1.3\%$ at $1.3\;$days after the burst with the Multiple Mirror Telescope \citep{Bersier2003}, but the polarization level observed with Very Large Telescope (VLT) is about $1-2\%$ at the same time epoch \citep{Masetti2003}.}. Suppose that the maximum level of observed PDs is $\sim 5\%$, then \Figref{fig:xi-dependence-peakPD} suggests that the large-scale model is consistent with observations if $\xi^2 \gtrsim 0.4$ and $f_B=1.0$. The microscopic-scale model is also consistent with observations if $\xi^2 \gtrsim 0.4$. \par

\textbf{GRB 091018}: The optical polarimetry for $0.13-2.3\;$days after the burst was obtained with VLT for GRB 091018 \citep{Wiersema2012}. The PA seems to show a mild $\deg{90}$ flip around $T\sim T_{\rm jb}$, but PD and PA change randomly at $T > T_{\rm jb}$. This behavior can be interpreted by the mildly anisotropic field, $\xi^2 \sim 0.4$, with the on-axis observer.\par

\textbf{GRB 121024A}: The optical afterglow with VLT for GRB 121024A was observed from $\sim 2.7$ hours to $\sim 1.2\;$days after the burst \citep{Wiersema2014}. This event clearly indicates a $\deg{90}$ flip around $T\sim T_{\rm jb}$, which can be interpreted by the highly or mildly anisotropic field, $\xi^2 < 0.5$, with the on-axis observer.\par

\textbf{GRB 020813}: The relatively densely sampled optical polarimetric datasets were obtained for GRB 020813 before and after $T_{\rm jb}$ with VLT \citep{Gorosabel2004}. This event shows no $\deg{90}$ flip around the jet break time ($0.4 - 0.9\;$days after the burst) but constant or slow change of PA. This can be explained in the case of the nearly isotropic-field, $\xi^2 \sim 0.8$. \cite{Lazzati2004} attempt to fit this event in the microscopic-scale field model with the structured jet, but they conclude that the contribution of the large-scale field is required to reproduce the observational data.\par

\textbf{GRB 191221B}: The first simultaneous optical and radio polarimetric data was obtained for GRB 191221B at $2.5\;$days after the burst with VLT and ALMA \citep[][see \Secref{subsec:obs-radio-afterglow} for the radio polarization]{Urata2023}. The optical polarization is detected at three epochs around $T\sim T_{\rm jb}$, which shows the almost constant PD and PA. Such temporal evolution of PDs and PAs is inconsistent with that of the microscopic-scale model. The sparse data points without substantial changes of PD and PA could be reproduced by the nearly isotropic-field $\xi^2 \sim 0.8$ in the large-scale model. In the microscopic-scale model, the contribution of the ordered field may reproduce the observed PD and PA \citep{Teboul&Shaviv2021}.\par

\textbf{Flip ratio:}
We also compare \Figref{fig:xi-dependence-90flip} and the observed PAs. According to the four optical polarimetric observations conducted around $T\sim T_{\rm jb}$, one event shows clear $\deg{90}$ flip (GRB 121024A), one event shows milder $\deg{90}$ flip (GRB 091018), and remaining two events show no flip (GRB 020813, GRB 191221B), i.e., the observed flip ratio is $\sim 25\%$. It could be interpreted by the microscopic-scale model if some of GRB jets are uniform while the others are structured. In the large-scale model, $\xi^2 < 0.1$ is suggested if all events have the same $\xi^2$. However, the more polarimetric observations over $T\sim T_{\rm jb}$ are necessary to put restrictions on the value of $\xi^2$.\par

\begin{deluxetable*}{l|l|c|c}[t]
    \tablecaption{The interpretation of afterglow polarimetric data with the large-scale field model and microscopic-scale field model.\label{tab:obs-summary}}
    \centering
    \tablehead{
        \colhead{observational feature} & 
        \colhead{GRB name} & \colhead{$\xi^2$ in large-scale model ($f_B=1$)} & \colhead{microscopic-scale model}
    }
    \startdata
        late time afterglow & Typical optical afterglow & $\gtrsim 0.4$ & \large{$+$} \\
         & GRB 091018          & $\sim 0.4$ & \large{$-$} \\
         & GRB 121024A         & $<0.5$ & \large{$+$} \\        
         & GRB 020813          & $\sim 0.8$ & \large{$-$} \\
         & GRB 191221B         & $\sim 0.8$ & \large{$-$} \\
        \hline
        early time afterglow & GRB 091208B & (RS) & (RS)\\
         & GRB 180720B         & $\sim 0.8$ & \large{$+$} \\
         & GRB 191016A & (RS) & (RS)\\
         & GRB 210610B         & $\sim 0.5$ & \large{$-$} \\
        \hline
        GW-associated event & GRB 170817A & $\gtrsim 0.7$ & \large{$+$} \\
        \hline
        radio polarization & GRB 171205A         & (Faraday effect) & \large{$-$} \\
                           & GRB 191221B         & $\sim 0.8$ & \large{$-$} \\
    \enddata
    \tablecomments{The third column shows $\xi^2$ for explaining the observed polarimetric features in the large-scale model. In the fourth column, "$+$/$-$" marks indicate that the observed polarimetric features can/cannot be explained by only the microscopic-scale turbulent field.}
\end{deluxetable*}

\subsection{observation of early afterglows}\label{subsec:obs-early-afterglow}
\textbf{GRB 091208B}: A high PD at early-phase forward-shock afterglow, PD $\sim10\%$, was detected for GRB 091208B with the Kanata Telescope at 6.0 minutes after the burst \citep{Uehara2012}. Such a high PD at $T \ll T_{\rm jb}$ is difficult to explain both in the large-scale and microscopic-scale models. It may be caused by contribution from the reverse shock emission \citep{Jordana-Mitjans2021}.\par

\textbf{GRB 180720B}: The optical polarization measurements were conducted for GRB 180720B with the Kanata Telescope during 70-20,000 sec after the burst, which covered both the reverse and forward shock dominated phases \citep{Arimoto2023:2310.04144v1}. The $\deg{90}$ flip with some fluctuations occurs during the transition from the reverse shock to the forward shock. \cite{Arimoto2023:2310.04144v1} suggests that large-scale toroidal and microscopic-scale turbulent magnetic field with $\xi^2>1$ in the reverse and forward shock, respectively, can account for the observed PD and PA. We suggest that large-scale model with $\xi^2 \sim 0.8$ can also explain the forward shock polarization in this event.\par

\textbf{GRB 191016A}: The simultaneous optical polarization in the V, R, and I bands for GRB 191016A was measured by ground based Liverpool Telescope from 3987-7687 sec after the burst \citep{Shrestha2022}. This event shows high PD $= 14.6 \pm 7.2 \%$ at the start of the flattening phase of the optical light curve. \cite{Shrestha2022} suggests a refreshed shock scenario in which the ejecta from the central engine has ordered field to interpret this high optical PD. Such a high PD in the early optical afterglow is difficult to explain both in the large and microscopic-scale models at the forward shock region.\par

\textbf{GRB 210610B}: The optical polarization for GRB 210610B was measured at three epochs, an early plateau phase at $0.12\;$days after the burst by Calar Alto Telescope, a phase of light curve break at $\sim 0.25\;$days and a light curve steepening phase at $1.3\;$days by VLT \citep{Fernandez2024}. It shows that initial PD $\sim 4\%$ during the early plateau phase goes to zero at the time when the light curve breaks, and then increases again to $\sim 2\%$ after the light curve steepened. The PA between the plateau phase and the steepened phase changes of $\deg{54}\pm \deg{9}$. It is difficult to explain this polarimetric feature with the microscopic-scale model because this model shows $\deg{90}$ flip or constant PA. On the other hand, the observed polarization is consistent with ordered field in a refreshed shock or/and large-scale model with $\xi^2 \sim 0.5$ in a forward shock.\par

\subsection{observation of GW-associated GRB (off-axis)}\label{subsec:obs-GW-asscociated-GRB}
\textbf{GRB 170817A}: The polarimetric observation for GW 170817/GRB 170817A, which was an off-axis GRB, was conducted by VLA at 2.8 GHz ($\nu>\nu_m$). The afterglow linear polarization of this event shows an upper limit of ${\rm PD} < 12\%$ (99\% confidence) at $T \approx 244\;{\rm day}$ \citep{Corsi2018}. \cite{Gill2020} shows that the microscopic-scale model with the marginally isotropic field is consistent with this upper limit. In our large-scale model, $\xi^2 \gtrsim 0.7$ is favored to explain this upper limit. In the future polarimetric observations of the off-axis GRB afterglow, we would distinguish the large-scale model from the microscopic-scale model by the random behavior of PDs and PAs if $\xi^2 \gtrsim 0.5$. However, if $\xi^2 \lesssim 0.5$, the behavior of polarization in the microscopic-scale model is similar to that in the large-scale model, then it is difficult to distinguish them.\par

\subsection{observation of radio afterglows}\label{subsec:obs-radio-afterglow}
\textbf{GRB 171205A}: The first detection of radio afterglow polarization was conducted for GRB 171205A by ALMA at $5.2\;$days after the burst \citep{Urata2019}. It shows the sub percent level of ${\rm PD} = 0.27\pm 0.04\%$ at 97.5\;GHz. The PA spectrum shows random behavior and PA varies by at least $\deg{20}$ in the millimeter range. This is inconsistent with our large-scale model because in this model PA does not show the frequency dependence of polarization within the radio band (c.f. \Figref{fig:PDPA-spectrum-model4} and \ref{fig:PDPA-spectrum-modelI1}). The microscopic model is also inconsistent for the same reason. However, when the Faraday effect due to non-heated electrons works in the large-scale model, the frequency dependence of PA can be explained \citep{Toma2008}.\par
\textbf{GRB 191221B}: As mentioned in \Secref{subsec:obs-late-afterglow}, GRB 191221B is the first simultaneous optical and radio polarimetric observations of GRB afterglow \citep{Urata2023}. This event shows that the PA difference between the optical and the radio is neither $\deg{0}$ nor $\deg{90}$, which is not consistent with the microscopic-scale field model but consistent with the large-scale field model. As argued in \Secref{subsec:obs-late-afterglow}, the nearly isotropic field $\xi^2 \sim 0.8$ is favored in the large-scale model from the constant optical PD and PA in time at $T\sim T_{\rm jb}$.\par

\section{Summary}\label{sec:summary}
The origin of strong magnetic field in the GRB forward-shock emission region is a long-standing problem. Polarimetric observations are powerful probes for the magnetic field structure in the shocked region, which is closely related to its amplification mechanism. So far, many theoretical works focus on the microscopic-scale turbulence, whose coherence length is much smaller than the thickness of the blast wave. Recently, K23 constructed the forward-shock polarization model with the large-scale turbulence, which is comparable to the thickness of the blastwave. In this work, we have investigated the dependence of the polarization in large-scale model on the anisotropy of magnetic field $\xi^2$, the observer's viewing angle $\theta_v$, and the ratio of ordered-field energy to fluctuated-field energy $\eta$. We summarize our results below:
\begin{enumerate}[(I)]
    \item In the case of the highly anisotropic field ($\xi^2<0.4,~\xi^2>2$) and the on-axis observer ($0<\theta_v<\theta_j$), the optical PDs show double peak and the optical PAs show $\deg{90}$ flip around the jet break time. This behavior is similar to that of the microscopic-scale model, and it is due to the similar local PA distribution at $\theta \sim 1/\gamma_l$. In the isotropic field case ($\xi^2 \sim 1$), however, both PD and PA change randomly and continuously in time (K23).
    
    \item We investigate the $\xi^2$ dependence of the number ratio of events which have PA $\deg{90}$ flips (`flip-ratio') and the maximum level of PD. In the large-scale model, the flip ratio decreases with $\xi^2$ and is $\sim 0\%$ for $\xi^2\gtrsim 0.5$. In the microscopic-scale model, the flip-ratio is $100\%$ for the uniform jets \citep{ST2021} and $0\%$ with the structured jets \citep{Rossi2004}, both of which are not consistent with the observational data.
    
    \item In the case of the off-axis observer and $\xi^2=1$, PDs and PAs show random behavior similar to that in the on-axis observer but realization-averaged PDs are about two times higher than the on-axis case. If the observer is off-axis and $\xi^2=0$, PDs take peak $\sim 40\%$ and PAs are almost constant, which is similar behavior to the microscopic-scale model \citep{Rossi2004}.
    
    \item If not only the fluctuated-field component but the ordered-field component is considered, the effect of the ordered field on polarization appears when $\eta \gtrsim 1$ in the optical band and $\eta \gtrsim 3$ in the radio band. The PD and PA are constant in time only when $\eta \gtrsim 50$. However, in this case, PD is much higher than the observed values.
\end{enumerate}
\par

The above results suggest that the large-scale model could account for all the polarimetric data that seem to be forward shock emission, including several events that are inconsistent with the microscopic-scale model. \Tabref{tab:obs-summary} suggests that $\xi^2$ in the large-scale model varies from event to event. Future observations will put more strict constraints on the turbulence model. We have found that three observational strategies are crucial. First, the increase of dense optical polarimetric observations around $T\sim T_{\rm jb}$ can constrain the flip ratio (\Secref{sec:anisotropic-field}). Second, the more simultaneous optical and radio polarimetric observation would catch epochs of radio PDs higher than the optical ones when $\xi^2 \sim 1$ (\Secref{sec:PD-spectrum}). Third, the early-time observation of off-axis events would constrain the coherence length scale and the field anisotropy of turbulence field (\Secref{sec:off-axis}). These observational approaches can help us constrain the configuration of the turbulent magnetic field and clarify the energy dissipation process at relativistic collisionless shocks.


\begin{acknowledgements}
We thank M. Saito for useful discussion. We utilized Science Lounge of FRIS CoRE for discussions many times. Numerical computations were in part performed on Draco, a computer cluster of FRIS, and in part carried out on Cray XC50 at Center for Computational Astrophysics, National Astronomical Observatory of Japan. This work is partly supported by Graduate Program on Physics for the Universe (GP-PU), Tohoku University (A.K.), by JST SPRING, Grant Number JPMJSP2114 (A.K.), by Grant-in-Aid for JSPS Fellows No. 24KJ0034 (S.T.), and by KAKENHI grant No. 24K00677 (J.S.).
\end{acknowledgements}

\appendix
\setcounter{figure}{0}
\renewcommand{\thetable}{\Alph{section}.\arabic{table}}
\renewcommand{\thefigure}{\Alph{section}.\arabic{figure}}

\section{The other realizations}\label{app:realization}
\begin{figure}[htbp]
    \centering
    \includegraphics[width=0.32\linewidth]{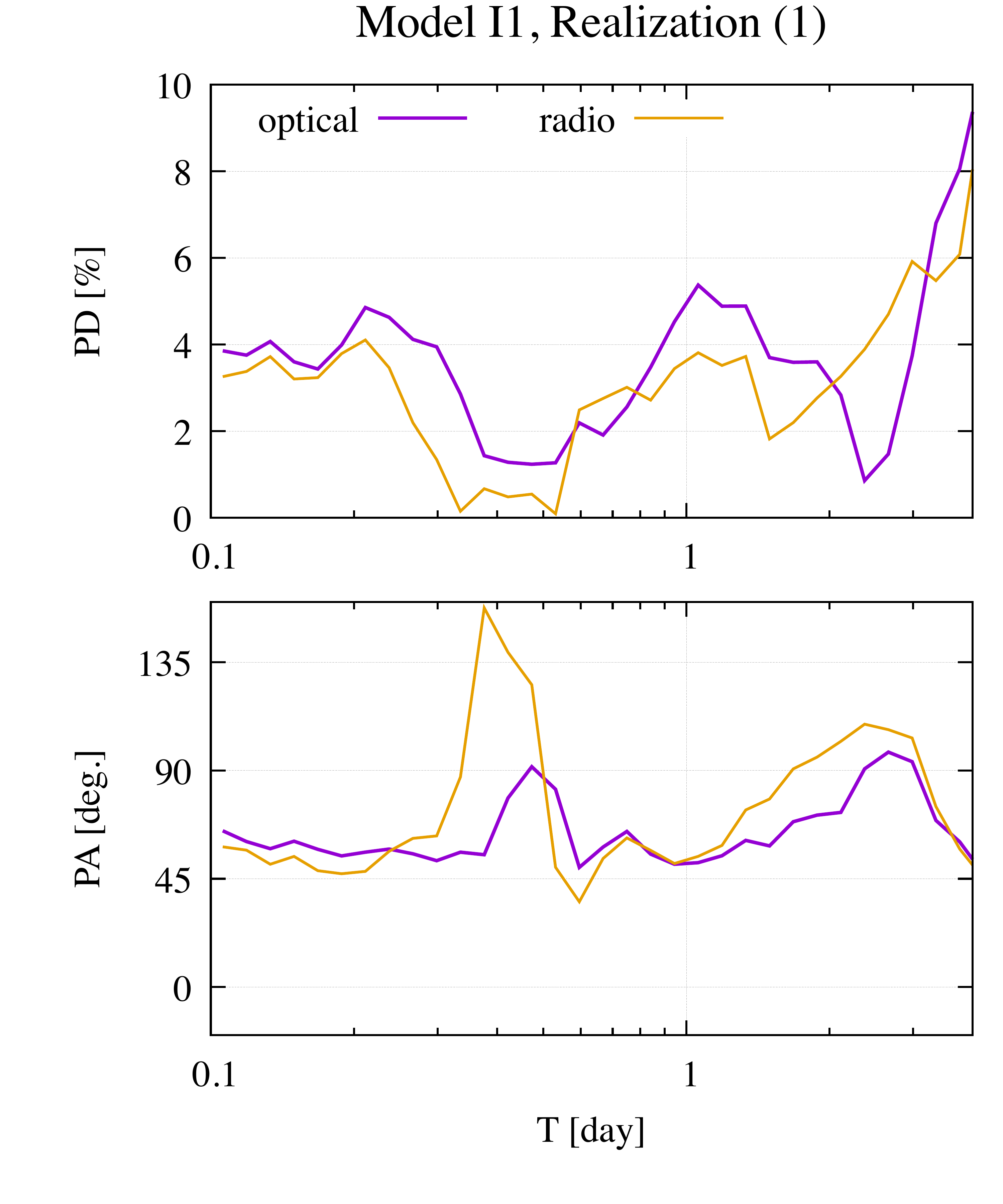}
    \includegraphics[width=0.32\linewidth]{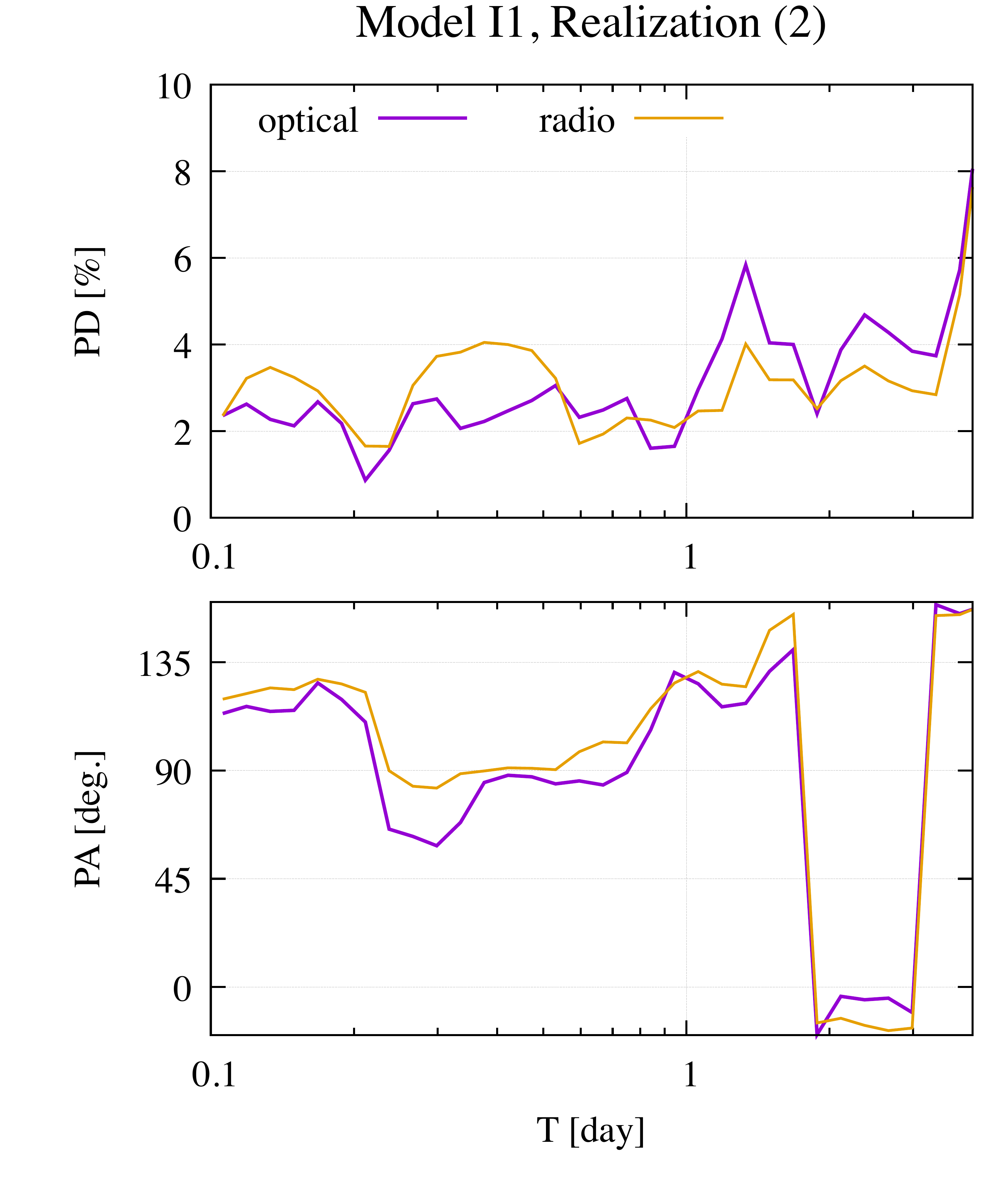}
    \includegraphics[width=0.32\linewidth]{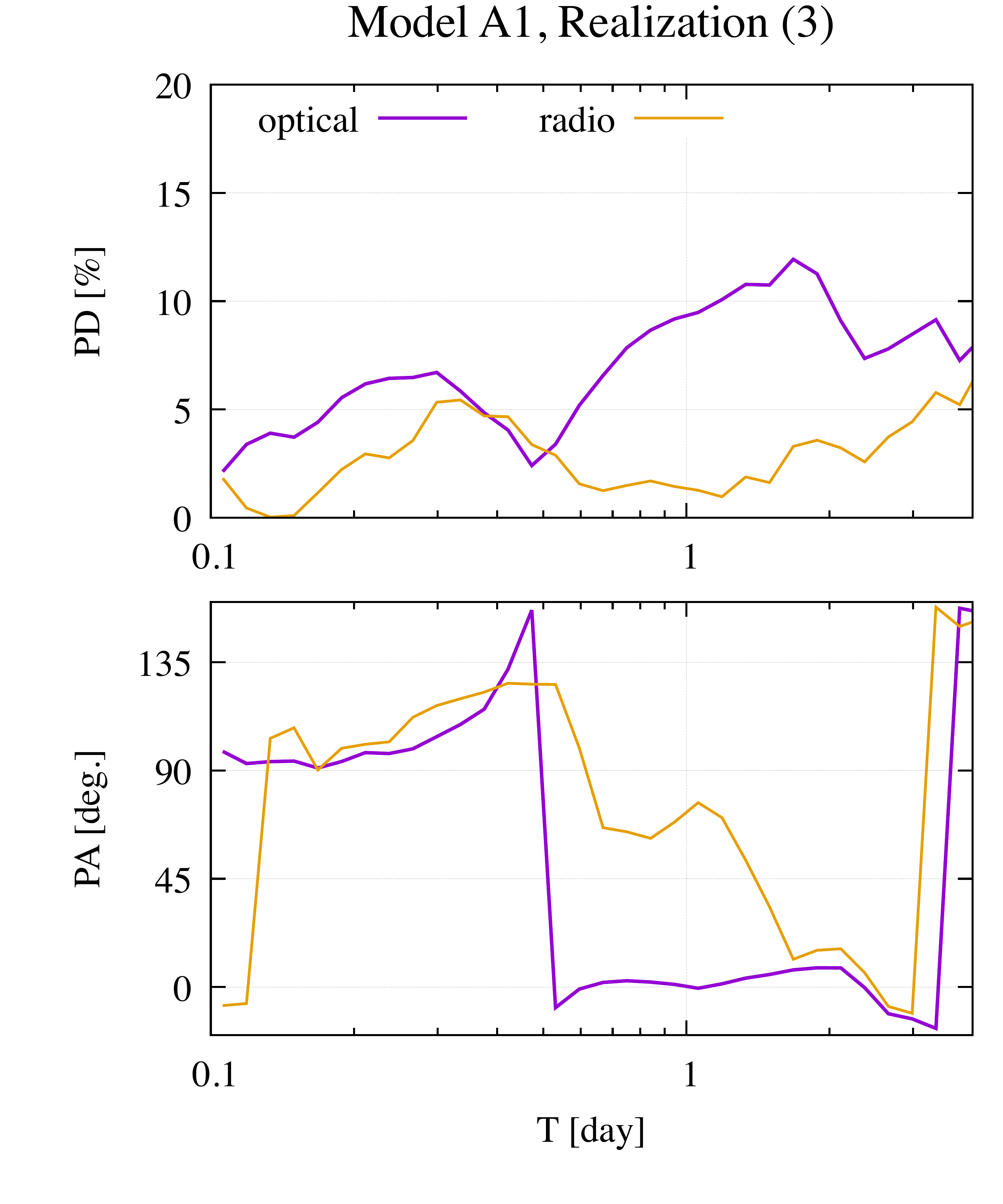}\\
    \includegraphics[width=0.32\linewidth]{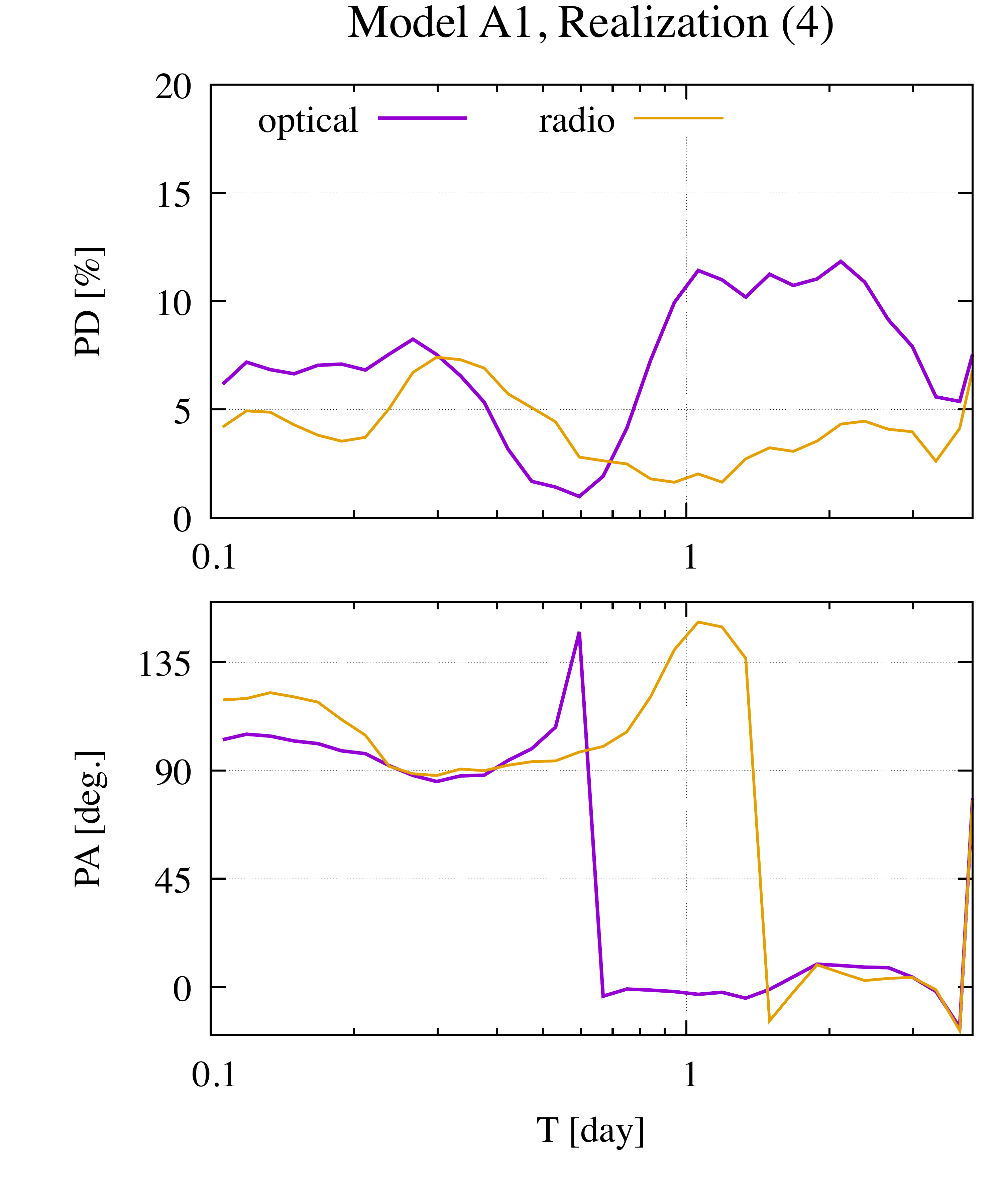}
    \includegraphics[width=0.32\linewidth]{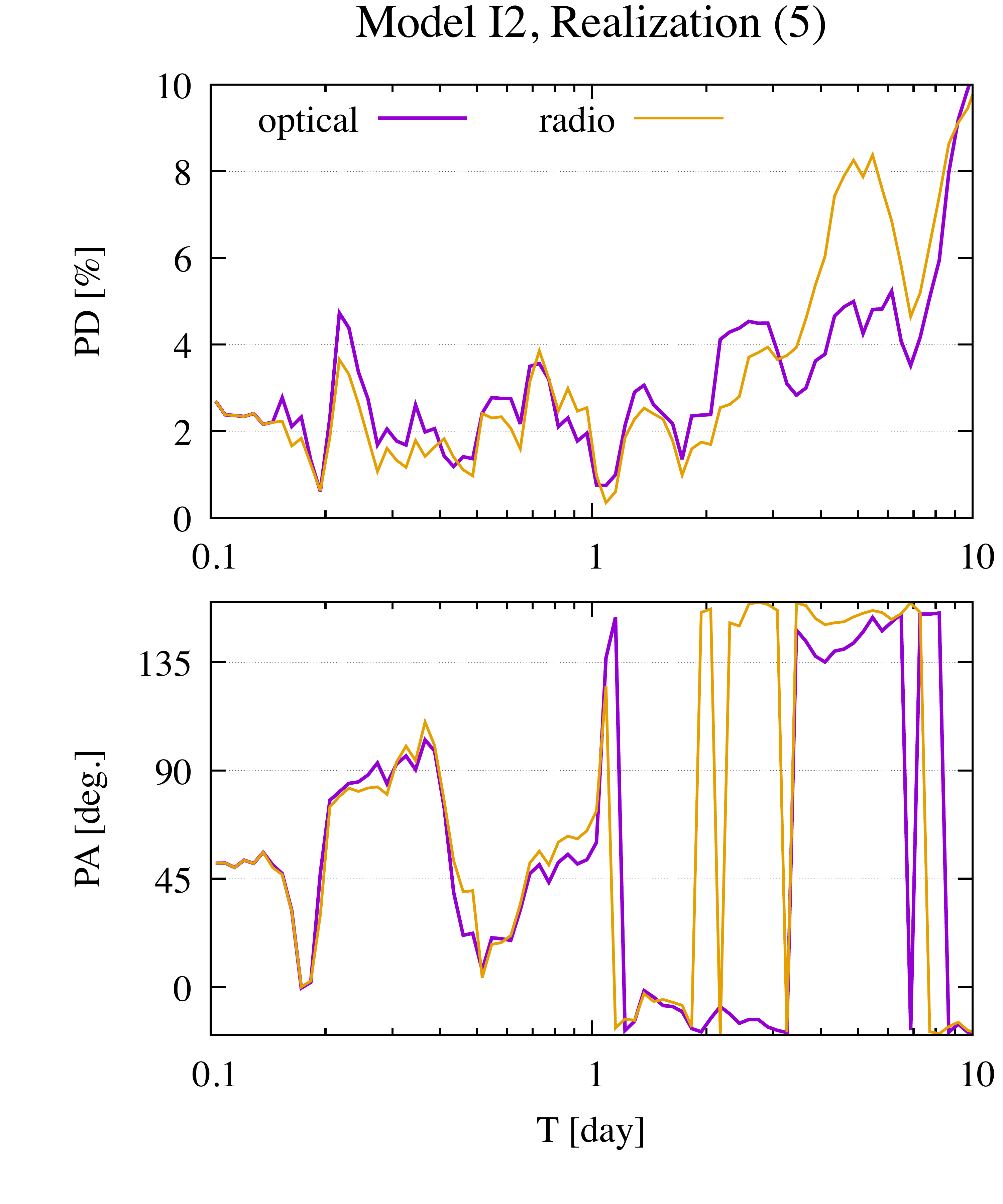}
    \includegraphics[width=0.32\linewidth]{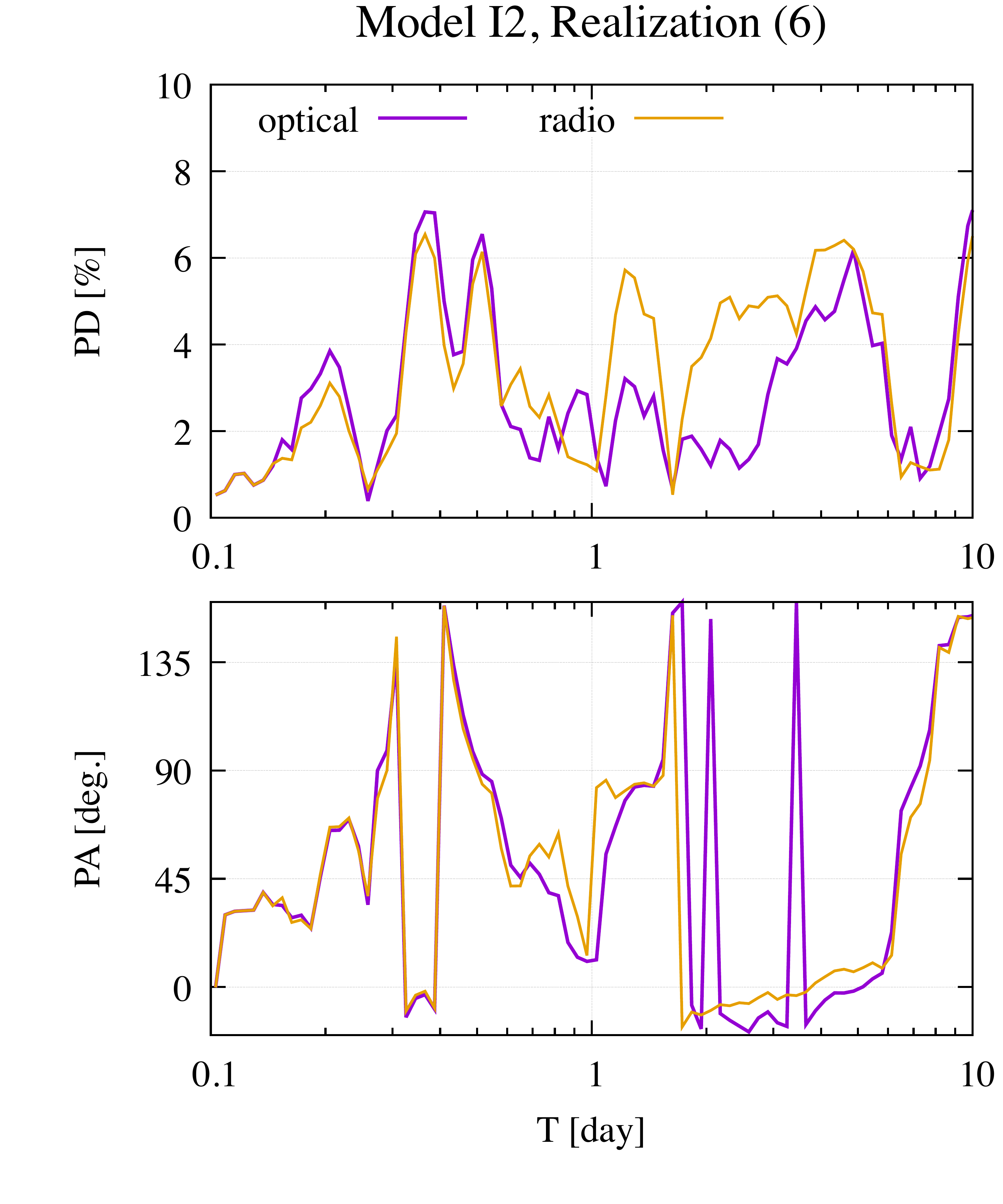}
    \caption{
    PDs and PAs as functions of time $T$ for Model I1, A1, I2, A2, and O at the frequencies $\nu_{\rm opt}$ (purple lines) and $\nu_{\rm radio}$ (orange lines) during $T = 0.1-4.0\;{\rm days}$. 
    Realizations (1)-(10) are the calculation results for different realizations of turbulent magnetic field from \Figref{fig:PDPA-curve-model1}, \ref{fig:PDPA-curve-model4}, \ref{fig:PDPA-curve-modelI2}, \ref{fig:PDPA-curve-modelA2}, \ref{fig:PDPA-curve-modelO}. The top and bottom panels represent the PD curves and the PA curves, respectively. 
    \label{fig:realization}}
\end{figure}

We show the PD and PA curves in other turbulent realizations for Model I1, A1, I2, A2, and O in \Figref{fig:realization}. From these figures, one can see that the explanations given in the main text are valid for different realizations from those given in the text.

\begin{figure*}[htbp]
    \centering
    \includegraphics[width=0.32\linewidth]{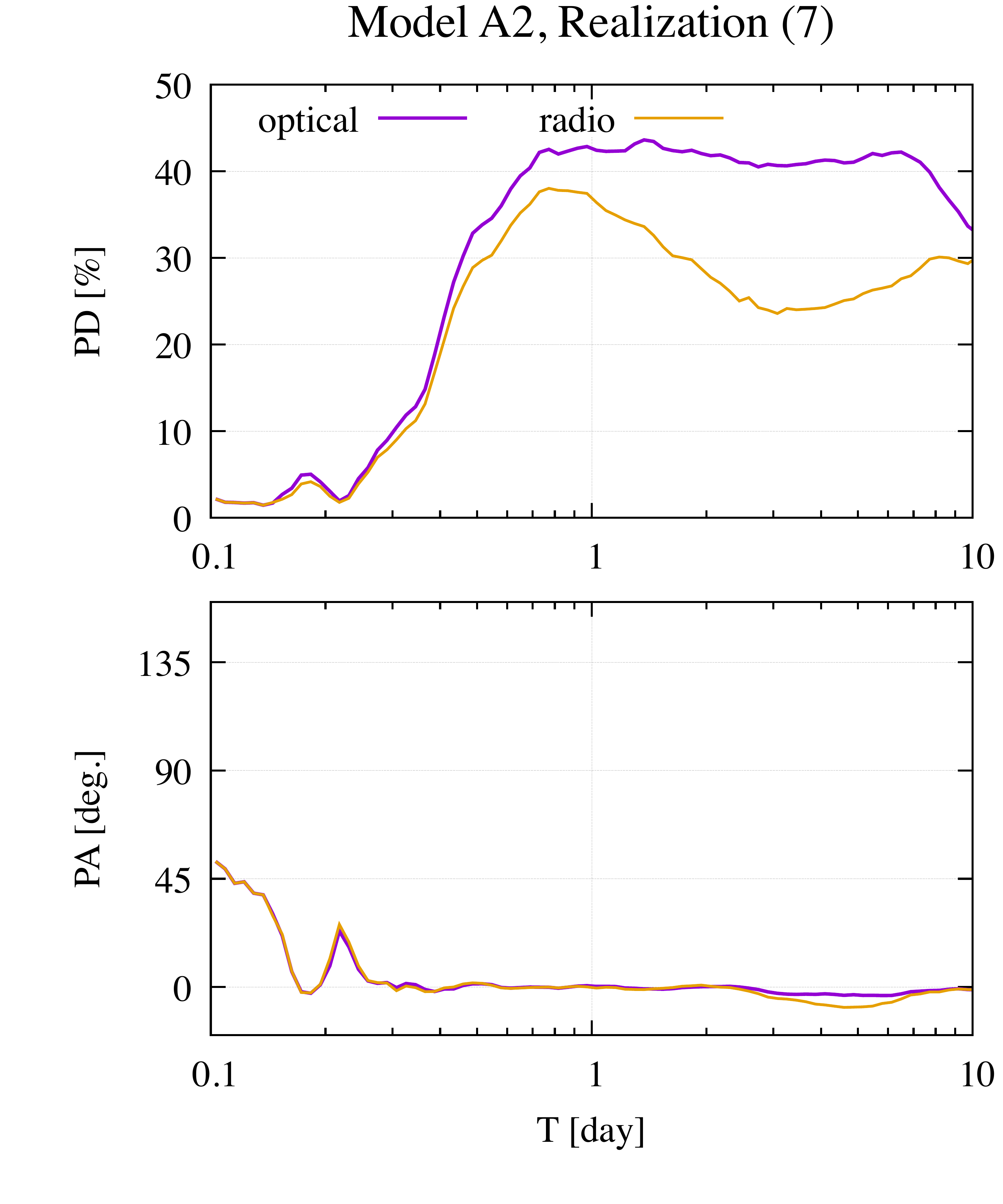}
    \includegraphics[width=0.32\linewidth]{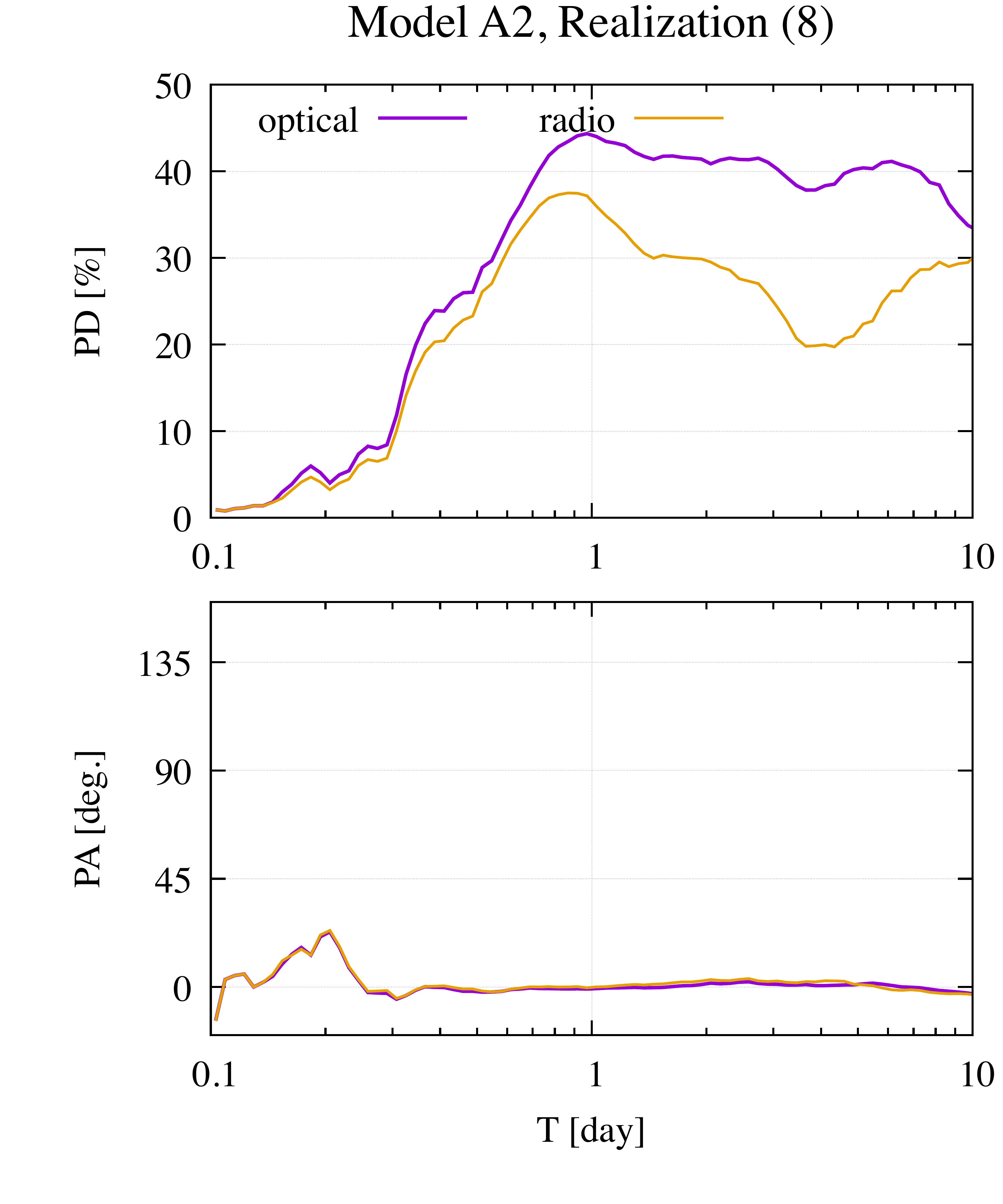}\\
    \includegraphics[width=0.32\linewidth]{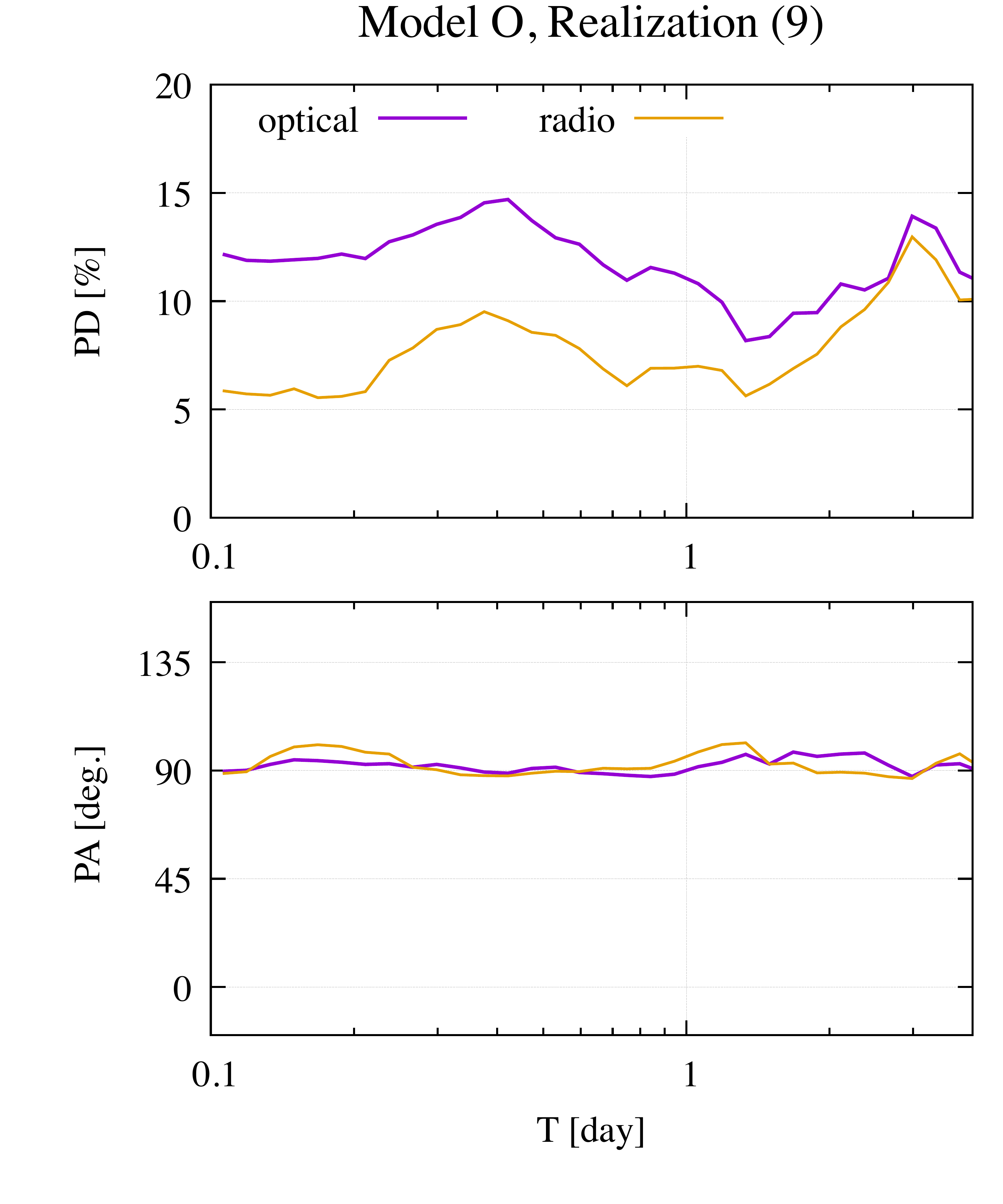}
    \includegraphics[width=0.32\linewidth]{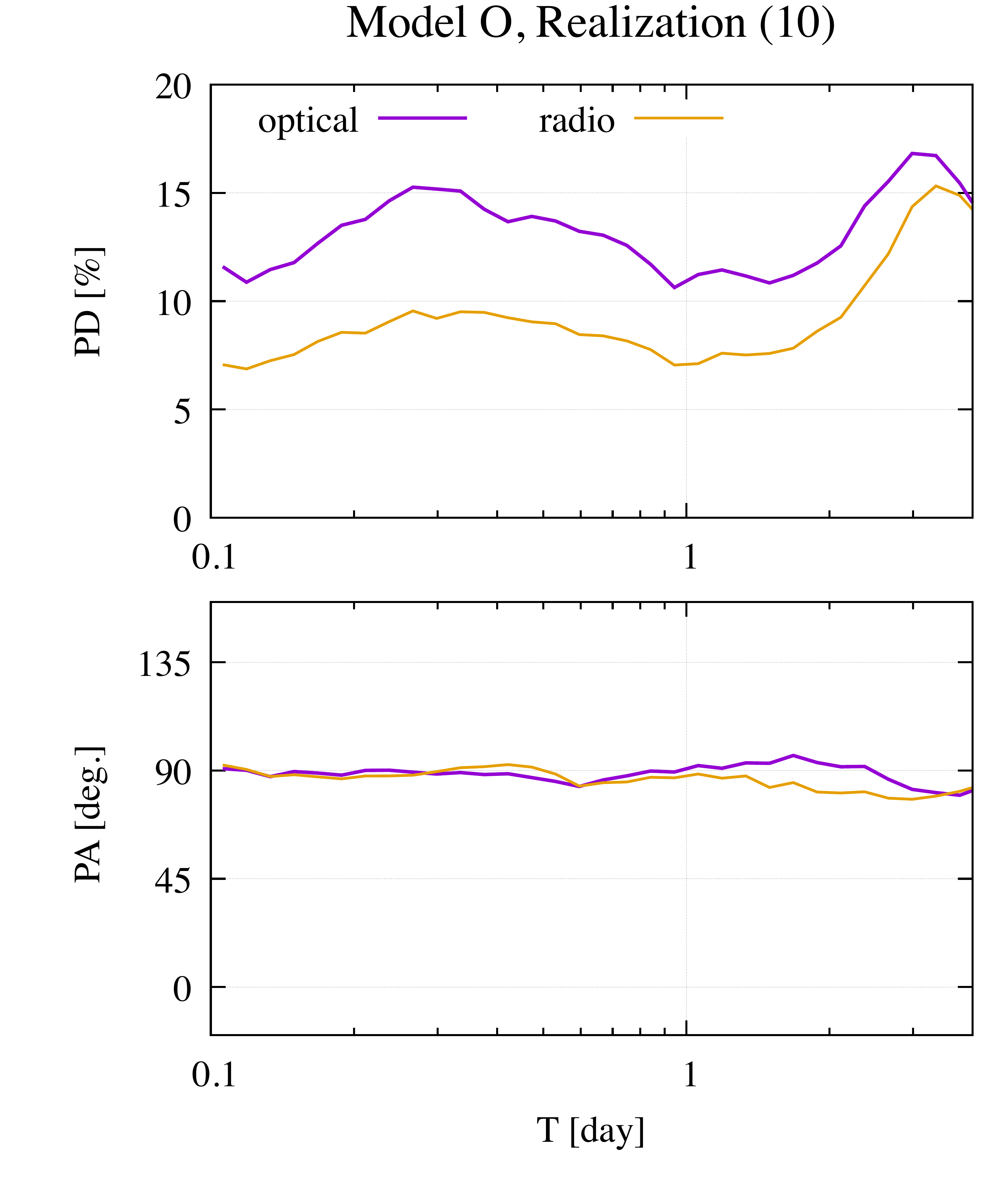}
    \par\vskip1ex
    {continue \Figref{fig:realization}}
     \par\vskip1ex
\end{figure*}

\bibliography{ms}{}
\bibliographystyle{aasjournal}
\end{document}